\documentclass[10pt]{iopart}
\usepackage{graphicx} 
\usepackage{amsmath}
\usepackage{xcolor}
\usepackage{caption}
\usepackage{subcaption}
\usepackage{hyperref}

\definecolor{background}{rgb}{0.97, 0.97, 0.97}

\usepackage[frozencache,cachedir=.]{minted}
\setminted[python]{bgcolor = background, samepage = True}
\setmintedinline[python]{bgcolor = background}

\begin{document}

\title{Simple Python tools for modelling few-level atom-light interactions}
\author{Lucy Downes$^1$}
\address{$^1$ Department of Physics, Durham University, South Road, Durham. DH1 3LE, UK}

\ead{
lucy.downes@durham.ac.uk}
\vspace{10pt}
\begin{indented}
\item[\today]
\end{indented}

\begin{abstract}
    Understanding the interactions between atoms and light is at the heart of atomic physics. 
    Being able to `experiment' with various system parameters, produce plots of the results and interpret these is very useful, especially for those new to the field. 
    This tutorial aims to provide an introduction to the equations governing near-resonant atom-light interactions and present examples of setting up and solving these equations in Python. Emphasis is placed on clarity and understanding by showing code snippets alongside relevant equations, and as such it is suitable for those without an excellent working knowledge of Python or the underlying physics. Hopefully the methods presented here can form the foundations on which more complex models and simulations can be built. All functions presented here and example codes can be found on GitHub. 
\end{abstract}

\section{Introduction}


The interactions of hydrogen-like atoms with narrowband near-resonant fields form the foundations of much of modern atomic physics.  
Being able to describe and model these interactions is therefore of great importance from both a theoretical and experimental perspective. 
The simplest way to describe atom-light interactions is to consider a classical picture, such as Einstein's rate equations.  
However classical descriptions do not give rise to many phenomena of interest in atom-light systems such as electromagnetically induced transparency (EIT)~\cite{Finkelstein2023}. In order to see how these effects emerge, a semi-classical approach must be used that takes into account aspects of the quantum mechanical nature of the system. 
The equations that this approach gives rise to cannot always be solved analytically, so numerical methods need to be applied. Even when analytic solutions can be found they often require making assumptions about the system so care needs to be taken to ensure these are valid. One way of exploring a system is by building a simple computational model and changing the system parameters to understand their effects. Having tools with which to build these simulations is an invaluable resource to both students and established scientists alike. 

Python is a high-level programming language that has proved to be a useful tool in scientific computing due to its ease of use, vast scope and free availability. 
Specialised open-source atomic physics packages are commonplace, allowing users to perform complex calculations such as model the electric susceptibility of an atomic gas~\cite{elecsus1, elecsus2} or integrate wavefunctions and calculate interaction potentials between highly-excited Rydberg atoms~\cite{Weber2017, Sibalic17, Robertson2021}. 
There are already a multitude of different Python packages designed to deal with solving problems involving interactions between atoms and light, for example QuTiP~\cite{Johansson2013}, mbsolve~\cite{Riesch2021}, PyLCP~\cite{Eckel2022} and MaxwellBloch~\cite{maxwellbloch}. Many focus on time-dependent Hamiltonians and solutions and as such are highly complex. Often they require installation of a package which provides `black box' solutions - functions are called `blind' without knowledge of the underlying equations or methods. This means it is difficult to understand how the solutions are reached and what assumptions are being made, and it is even harder to troubleshoot when things go wrong.

Here we present a simple way of exploring the steady state behaviour of few-level atom-light systems using basic Python functions, without the need to install multiple packages beyond the standard Python libraries. \mintinline{python}{NumPy} (numerical Python~\cite{numpy20}), \mintinline{python}{SciPy} (scientific Python~\cite{scipy20}) and \mintinline{python}{SymPy} (symbolic Python~\cite{sympy17}) are all that is required, alongside \mintinline{python}{Matplotib} for producing plots~\cite{Hunter07}.  
We will outline how all of the Python functions are created from the equations, how they are manipulated and then how the steady-state solution can be found. We will also present examples of using the defined functions to create plots and interpret the underlying physics for both zero and finite temperature systems. 

\subsection{Paper Structure}
We will first review the semi-classical approach to near-resonant atom-light interactions using the density matrix to describe an atom-light system. We will derive the optical Bloch equations (OBEs) and look at how measurable quantities can be extracted to compare to experiments. 
In the next section we will look at two possible solution methods and the regimes in which they are valid. These methods are presented alongside relevant code snippets to give some insight into how the equations are translated into Python functions. 
Given lists of the system parameters, we will use \mintinline{python}{Sympy} to derive the OBEs from an interaction Hamiltonian and decay/dephasing operator, then expresses these as matrix equations and uses a singular value decomposition to solve them. 
We then look at some examples of using the code to generate plots of quantities of common interest in experiment, and use these to understand some of the phenomena observed in atom-light systems. 
We finish by demonstrating how the functions could be used to account for effects such as Doppler broadening in systems with a finite temperature.

\subsection{Notes on code and examples}
In this tutorial, selected code snippets are presented to illustrate how the equations outlined are coded and solved. All functions outlined here are available in a module on GitHub (\url{https://github.com/LucyDownes/OBE_Python_Tools}) alongside Jupyter notebooks showing how the functions are created and some examples of their use. 
All codes used to generate the plots in this work are also available there. 
For most plots, parameters are in units of $2\pi\,\rm{MHz}$. The factor of $2\pi$ is omitted from plot axis labels for clarity. 
In examples, parameters were chosen such that a particular effect is clearly visible in the plot. Hence they are purely illustrative and not intended to represent any particular physical experiment. 
All of the code presented here was written in Python 3.8 but could very easily be made backwards compatible with Python 2.7. 
The code is in no way optimised for speed, it is intended to be as transparent as possible to aid insight. 
Variable names have been chosen to align with the notation used in this paper. Parameters (and hence variables) referring to a field coupling two levels $i$ and $j$ with have the subscript $ij$, whereas parameters that are relevant to a single atomic level $n$ will have the single subscript $n$. Variable names with upper/lower-case letters denote upper/lower-case Greek symbols respectively, for example \mintinline{python}{Gamma} denotes $\Gamma$ while \mintinline{python}{gamma} denotes $\gamma$.

\section{Theory}
\subsection{Semi-Classical Atom-Light Interactions}
In order to model how a single atom interacts with electromagnetic (EM) fields, we take a semi-classical approach in which the radiation field is considered as classical but the atom is considered quantum mechanically \cite{Cohen92, loudon2000}. 

The time evolution of the atomic wavefunction $|\psi\rangle$ is described by the time-dependent Schr\"{o}dinger equation
\begin{equation}
    i\hbar\frac{\partial}{\partial t}|\psi\rangle = H|\psi\rangle.
\end{equation}
The total Hamiltonian $H$ of the system is made up of two parts
\begin{equation}
H = H_0 + H_{I}(t),
\label{eqn:H}
\end{equation}
where $H_0$ is the bare atomic Hamiltonian in the absence of any radiation and $H_{I}$ describes the interaction with a time-dependent electromagnetic field. 
To begin with we will ignore the effects of dephasing processes and focus only on the coherent dynamics of the system. 

First let us consider a two level atom with non-degenerate levels $|1\rangle$ and  $|2\rangle$ having energies of $E_1 = \hbar\omega_1$ and $E_2 = \hbar\omega_2$. 
These are known as the bare states of the system, and form a complete basis set.
Any state of the atom $|\psi\rangle$ can then be written as a superposition of these bare states, which for our two-level atom can be written
\begin{equation}
    |\psi\rangle = c_1|1\rangle + c_2|2\rangle
    \label{eqn:state_vect}
\end{equation}
where the coefficients $c_{i}$ represent the probability amplitude of being in state $|i\rangle$. 

In the basis of bare states we can write the atomic Hamiltonian as
\begin{equation}
H_0 = \begin{pmatrix}
\hbar\omega_1 & 0 \\
0 & \hbar\omega_2
\end{pmatrix}. 
\end{equation}
The eigenvalues of this Hamiltonian give the energies of the two levels. 
By defining the transition frequency $\omega_0 = \omega_2 - \omega_1$, we can shift the zero point energy and rewrite the Hamiltonian as
\begin{equation}
H_0 = \begin{pmatrix}
0 & 0 \\
0 & \hbar\omega_0
\end{pmatrix}. 
\end{equation}
The energy difference between the states remains the same ($\hbar\omega_0$), we have just defined the energy of the lowest state to be zero. 

The levels are coupled by a laser field with frequency $\omega$. Since we are considering near-resonant excitation, we have that $\omega \approx\omega_0$, but we can quantify the difference between the laser frequency and the resonant frequency of the transition as $\Delta = \omega - \omega_0$. This is known as the detuning. 

The interaction Hamiltonian $H_I$ takes the form $-\mathbf{d}\cdot \mathbf{E}$ where $\mathbf{E}$ is the electric field and $\mathbf{d} = -e\mathbf{D}$ is the electric dipole operator.

Treating the electromagnetic field classically means it can be modelled as a simple oscillating field, meaning it can be written as $\mathbf{E} = E_0 \hat{\boldsymbol{e}}\cos(\omega t)$ where $E_0$ is the amplitude of the field, $\hat{\boldsymbol{e}}$ is a polarisation vector and $\omega$ is the frequency of the field. 
Here we have neglected the spatial dependence of the field, an approximation known as the dipole approximation.
This approximation is valid since the scale of the atomic wavefunction is much smaller than the wavelength of the applied fields, so the field can be considered uniform across the extent of the atom. 

The field interacts with the atom via the dipole operator. This can be written as
\begin{equation}
    \hat{\mathbf{D}} = \sum_{i,j} \mathbf{D}_{ij}|i\rangle\langle j| 
\end{equation}
where $|i\rangle$ and $|j\rangle$ are basis states of the bare atomic Hamiltonian. The term $\mathbf{D}_{ij} = \langle i|D|j\rangle$ is known as the matrix element \cite{Cohen92}. 
Now we can write the interaction Hamiltonian as
\begin{equation}
    H_I = eE_0 \cos(\omega t)\sum_{i,j} \mathbf{D}_{ij}\cdot \hat{\mathbf{e}}|i\rangle\langle j|.
\end{equation}
It is convenient at this point to define a new quantity 
\begin{equation}
    \Omega_{ij} = \frac{eE_0}{\hbar}\langle i|\mathbf{D}\cdot \hat{\mathbf{e}}|j\rangle
\end{equation}
which is known as the Rabi frequency. This defines the strength of the coupling between the levels $i$ and $j$ due to the field, and has units of frequency. It depends on both the intensity $E_0$ of the field and the polarisation through $\hat{\mathbf{e}}$. 
From the definition of the Rabi frequency we can see that $\Omega_{ii} = 0$, and $\Omega_{ji} = \Omega_{ij}^*$. 

Our interaction Hamiltonian now becomes
\begin{equation}
    H_I = \frac{\hbar\Omega}{2}\left(e^{i\omega t} + e^{-i\omega t}\right)\left[|1\rangle\langle 2| + |2\rangle\langle 1|\right],
\end{equation}
where we have dropped the subscript on the Rabi frequency since there is only one driving field. 
The terms in the round brackets describe the field, and can be thought of as describing the emission and absorption of a photon respectively. The operators in the square brackets describe the atom, and are related to a transition $|2\rangle\rightarrow |1\rangle$ and $|1\rangle\rightarrow |2\rangle$ (lowering and raising operators). 
The process whereby the atom makes the transition $|1\rangle\rightarrow |2\rangle$ after absorbing a photon, or makes the transition $|2\rangle\rightarrow |1\rangle$ by emitting a photon are more important than the other processes, so we will keep only the two terms that describe these \cite{Cohen92}. This is an approximation known as the `rotating wave approximation', and means the interaction Hamiltonian now has the form
\begin{equation}
    H_I = \frac{\hbar\Omega}{2}\left(e^{i\omega t}|1\rangle\langle 2| + e^{-i\omega t}|2\rangle\langle 1|\right).
\end{equation}
Combining this with the atomic Hamiltonian $H_0$, we get the total Hamiltonian
\begin{equation}
    H = \hbar \omega_0|2\rangle\langle 2| + \frac{\hbar\Omega}{2}\left(e^{i\omega t}|1\rangle\langle 2| + e^{-i\omega t}|2\rangle\langle 1|\right),
\end{equation}
which could also be expressed in matrix form as
\begin{equation}
    H = \frac{\hbar}{2}\begin{pmatrix}
        0 & \Omega e^{i\omega t} \\
        \Omega e^{-i\omega t} & 2\omega_0
    \end{pmatrix}.
\end{equation}

In order to get rid of the explicit time dependence in the Hamiltonian, we can perform a frame transformation into a rotating frame. In this simple 2-level case this can be done\footnote{More generally, this frame transformation is performed by the application of a unitary operator of the form $\hat{U} = e^{i\omega t |2\rangle\langle2|}$. The transformed state vector is then $|\Tilde{\psi}\rangle = \hat{U}|\psi\rangle$, and the Schr\"{o}dinger equation becomes 
\begin{equation*}
    i\hbar\frac{\partial}{\partial t}|\Tilde{\psi}\rangle = \Tilde{H}|\Tilde{\psi}\rangle
\end{equation*}
where the transformed Hamiltonian $\Tilde{H} = \hat{U}H\hat{U}^\dagger + i\hbar\dot{\hat{U}}\hat{U}^\dagger$.} by introducing new variables $\Tilde{c}_1(t)$ and $\Tilde{c}_2(t)$ and writing the state vector as 
\begin{equation}
    |\psi\rangle = \Tilde{c}_1|1\rangle + \Tilde{c}_2e^{-i\omega t}|2\rangle.
\end{equation}

By inserting this expression for the state vector and the total Hamiltonian into the time-dependent Schr\"{o}dinger equation, it is possible to find equations for the time dependence of the coefficients $\Tilde{c}_1(t)$ and $\Tilde{c}_2(t)$. This can also be written in the form of a time-dependent Schr\"{o}dinger equation with the new time-independent Hamiltonian
\begin{equation}
    \Tilde{H} = \frac{\hbar}{2}\begin{pmatrix}
        0 & \Omega \\
        \Omega & -2\Delta
    \end{pmatrix},
    \label{eqn:H_2lvl}
\end{equation}
where $\Delta = \omega - \omega_0$ is the detuning we defined earlier. 
This is the most commonly used form of the Hamiltonian, and the form we will use from here onwards.

This combined atom-light Hamiltonian allows us to explore properties of the coupled atom-light system.
The effective energy levels of the atom-light system $E_{\pm}$, sometimes called eigenenergies, are found as the eigenvalues of this Hamiltonian and are given by
\begin{equation}
E_{\pm} = -\frac{\hbar\Delta_{12}}{2}\pm\frac{\hbar}{2}\sqrt{{\Delta_{12}}^2+{\Omega}^2}.
\label{eqn:dressed_eigs}
\end{equation}
The eigenvectors of $H$ are called `eigenstates'.
In the absence of the field (i.e. for $\Omega = 0$ and $\omega = 0$) the eigenstates are the bare states of the system ($|1\rangle$ and $|2\rangle$) with eigenenergies of $0$ and $\hbar\omega_0$ respectively, 
but when considering the coupled atom-field system the bare states are no longer eigenstates.  
For a near resonant field (i.e. for $\Delta\approx 0$) the eigenstates are given by
\begin{equation}
|\pm\rangle = \frac{1}{\sqrt{2}}\left(|1\rangle \pm |2\rangle\right).
\end{equation}
These new states are known as dressed states, and are linear superpositions of the bare states. 
From the eigenenergies in equation~\ref{eqn:dressed_eigs} it can be seen that for increasing $\Omega$ the levels $E_{\pm}$ will be shifted further from each other; an effect known as the AC Stark effect or light shift \cite{Foot}.
\begin{figure}
    \centering
    \includegraphics[width=0.7\textwidth]{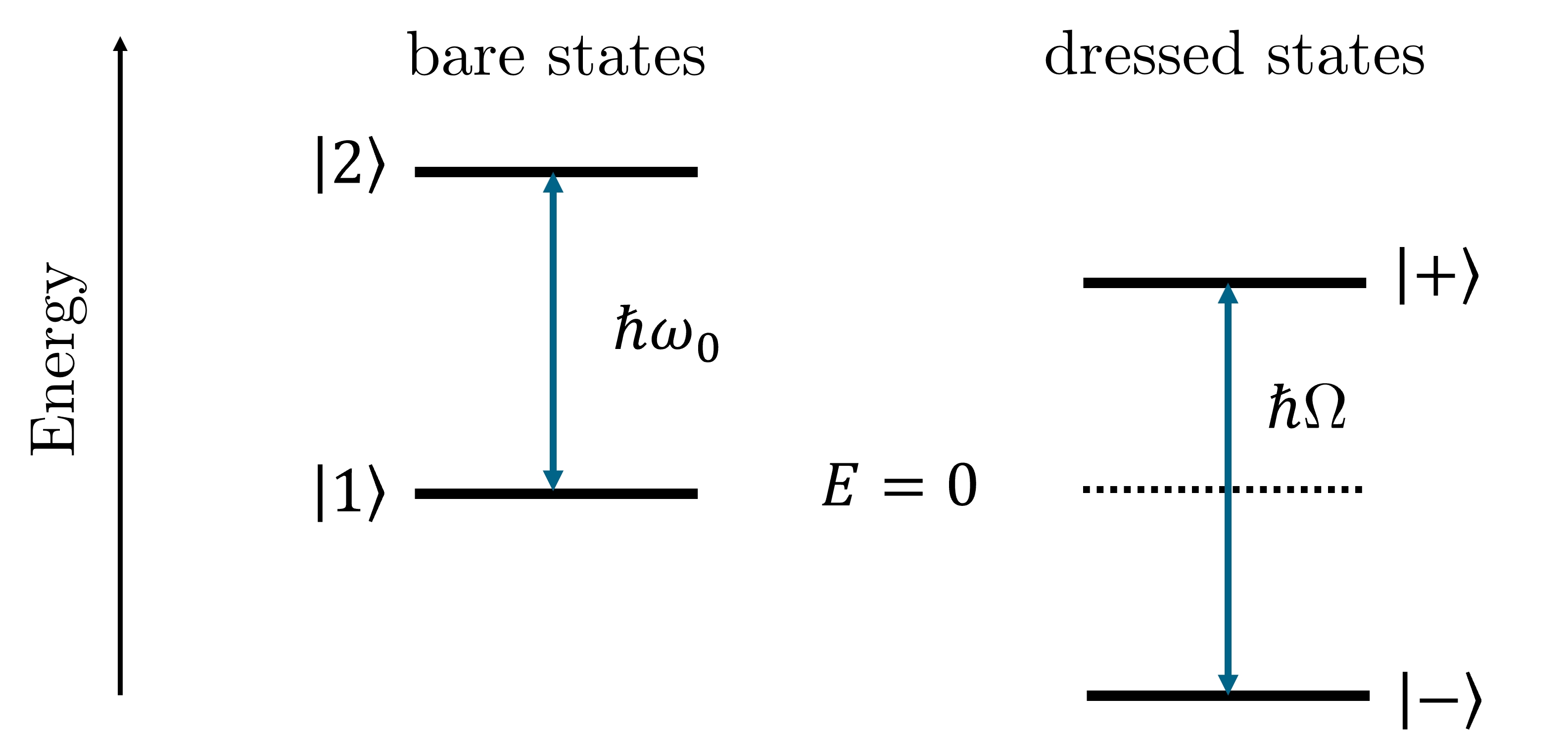}
    \caption{In the absence of any field coupling the atomic states, the bare states $|1\rangle$ and $|2\rangle$ are the eigenstates of the system, separated in energy by $\hbar\omega_0$. In the coupled atom-light system, the bare states are no longer eigenstates, instead we need to consider the dressed states $|\pm\rangle$, which have eigenenergies separated by $\hbar\Omega$ for a resonant field ($\Delta=0$).}
    \label{fig:2lvl_dressed_states}
\end{figure}

\subsection{Optical Bloch Equations}
\begin{figure}
    \centering
    \includegraphics[width=\textwidth]{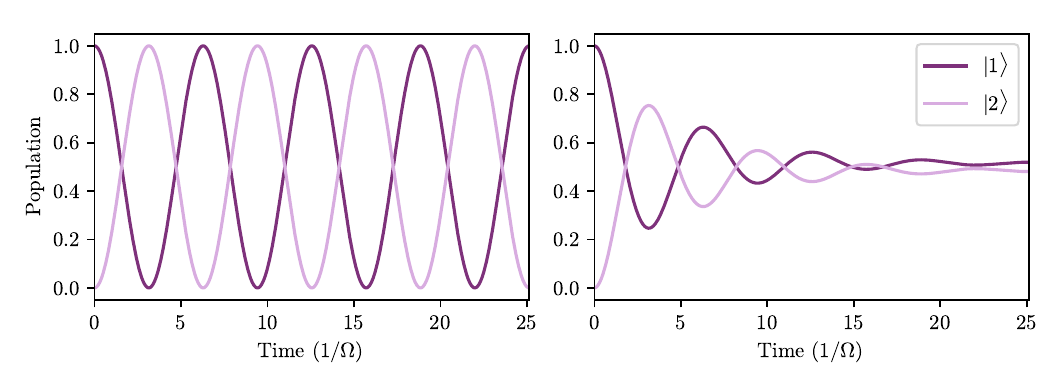}
    \caption{\textit{Left:} Evolution of the populations of both the ground (dark purple) and excited (light purple) state in a driven two-level system without decay. The populations oscillate with characteristic frequency $\Omega$, so-called Rabi oscillations. \textit{Right:} The same two-level system but including spontaneous decay from the excited state into the ground state at a rate $\Gamma_2 = \Omega/4$. The system quickly reaches the steady state at which point the population of both states is constant. The ratio of population in each state is related to the ratio between the decay rate and the Rabi frequency, as shown in Figure~\ref{fig:Time_Dep_2lvl_b}.}
    \label{fig:Time_Dep_2lvl_a}
\end{figure}

If we were to solve the Schr\"{o}dinger equation for the coherent dynamics of the state vector $|\psi\rangle$ as defined earlier, we would find that the solutions are oscillatory. This oscillation of population between the two states are known as Rabi oscillations, and occur at a frequency defined by the Rabi frequency. This behaviour can be seen in the left-hand panel of Figure~\ref{fig:Time_Dep_2lvl_a}. 
However in real atomic systems there are incoherent processes such as decay between the levels which need to be taken into account which is not possible by considering the state vector alone.
To this end we introduce the density operator defined as $\hat{\rho} = |\psi\rangle\langle\psi|$ and consider its evolution instead of that of the state vector. 

In the basis of bare states, the density operator can be written in matrix form. 
Using the definition of the state vector from equation~\ref{eqn:state_vect} we have 
\begin{equation}
    \hat{\rho} = \begin{pmatrix}
        |c_1|^2 & c_1 c_2^* \\
        c_2 c_1^* & |c_2|^2
    \end{pmatrix}.
\end{equation}
The diagonal elements of the density matrix describe the populations of the bare states $|1\rangle$ and $|2\rangle$, while the off-diagonal elements describe the coherence between states. 
It is worth noting that using the density matrix allows us to include more information than the state vector alone. For example it is possible to describe mixed states, states that are a statistical mixture of quantum states instead of a linear superposition, which cannot be described by a single state vector. For a pure state (one that can be described by a linear superposition of bare states) the trace of the square of the density matrix ($\mathrm{Tr}[\rho^2]$) is equal to 1, which is not the case for mixed states. 

The coherent time evolution of the density matrix is given by the Liouville von-Neumann equation \cite{Cohen92}
\begin{equation}
i\hbar\frac{\partial\hat{\rho}}{\partial t} = \left[\hat{H}, \hat{\rho}\right],
\end{equation}
in which $\left[\hat{H}, \hat{\rho}\right]$ represents the commutator of the total Hamiltonian $\hat{H}$ and the density operator $\hat{\rho}$. This equation is analogous to the time-dependent Schr\"{o}dinger equation for the state vector. 
In order to include dissipative effects we use the Master equation 
\begin{equation}
\frac{\partial\hat{\rho}}{\partial t} = -\frac{i}{\hbar}\left[\hat{H}, \hat{\rho}\right] + \hat{\mathcal{L}}(\hat{\rho}),
\label{eqn:Master}
\end{equation}
where again $\hat{H}$ is the Hamiltonian of the atom-light system and the Lindblad operator $\hat{\mathcal{L}}$ describes the decay/dephasing in the system.

The dissipative processes can be split into two separate parts; one part describing the spontaneous decay within the atom ($\hat{L}_{\rm{decay}}$) and another describing dephasing due to the finite linewidths of the driving fields ($\hat{L}_{\rm{dephasing}}$). The total operator will then be the sum of these parts 
\begin{equation}
\hat{\mathcal{L}} = \hat{L}_{\rm{decay}} + \hat{L}_{\rm{dephasing}}.
\end{equation}
First we focus on the operator $\hat{L}_{\rm{decay}}$ describing spontaneous decay between levels. The decay rate of population out of each level is given by $\Gamma_i$. For our two-level atom we assume that since the state $|1\rangle$ is the ground state, there is no decay out of this level. This amounts to saying $\Gamma_1 = 0$. We also assume that the system is closed; population remains within the two-level system and is not lost to an external reservoir. This means that any population that decays from $|2\rangle$ must end up back in $|1\rangle$. The amount of population transfer will depend on the population of the excited state. Hence the diagonal elements of the decay operator describing the change in the population of the ground and excited state will be given by $\Gamma_2\rho_{22}$ and $-\Gamma_2\rho_{22}$ respectively.  
Since decay can only ever cause a loss of coherence, the off-diagonal elements of the decay operator are always negative, and decay at the average rate of decay for the two levels. 
This means that for our 2-level system the decay operator can be written in matrix form as
\begin{equation}
    \hat{L}_{\rm{decay}} = \begin{pmatrix}
        \Gamma_2\rho_{22} & -\frac{\Gamma_2}{2}\rho_{12} \\
        -\frac{\Gamma_2}{2}\rho_{21} & - \Gamma_2\rho_{22}
    \end{pmatrix}.
\end{equation}
The action of decay is to damp the oscillations, meaning that eventually the system reaches an equilibrium, a so-called steady state. This is illustrated in the right-hand panel of Figure~\ref{fig:Time_Dep_2lvl_a}. In continuous-wave experiments, this steady state is reached in very short timescales compared to the measurement time, so the steady state solution is often of most interest.
\begin{figure}
    \centering
    \includegraphics[width=\textwidth]{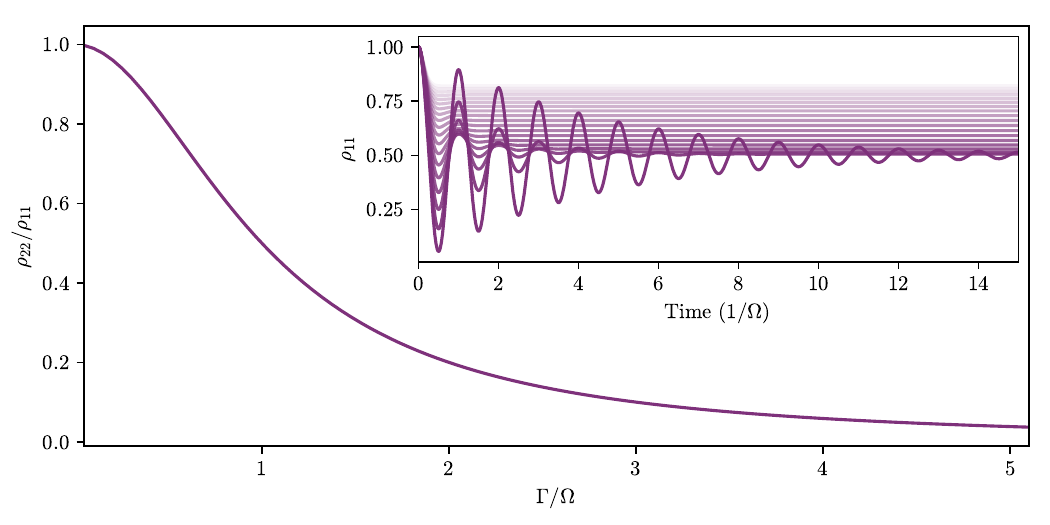}
    \caption{The ratio of excited to ground state population in the steady state for varying decay rates. For low values of $\Gamma_2$, the populations of the two states are equal in the steady state, whereas for increasing values of $\Gamma_2$, less and less population is transferred to the excited state. \textit{Inset:} Examples of the time evolution of the ground state population for values of $\Gamma_2$ between $\Omega/10$ and $2\Omega$ (dark to light lines). Oscillations are more strongly damped for larger values of $\Gamma_2$.}
    \label{fig:Time_Dep_2lvl_b}
\end{figure}

Secondly we establish the form of the dephasing operator $\hat{L}_{\rm{dephasing}}$ which accounts for the fact that the driving field can never be truly single-frequency, but will instead have a finite width. 
The dephasing due to the finite laser linewidths only affects the coherences (off-diagonal density matrix elements). 
Since we have only two levels, we only need to include effects due to a single field. In this case the dephasing operator takes the form
\begin{equation}
    \hat{L}_{\rm{dephasing}} = \begin{pmatrix}
        0 & -\gamma_{12}\rho_{12} \\
        -\gamma_{21}\rho_{21} & 0
    \end{pmatrix}
\end{equation}
where $\gamma_{12}=\gamma_{21}$ is the linewidth of the field coupling the levels $|1\rangle$ and $|2\rangle$.
Now we have expressions for the Hamiltonian and decay operator, we can substitute them into equation~\ref{eqn:Master} to get an expression for the time evolution of the density matrix. 
In the steady-state, the solution is not time-dependent. 
This steady-state solution can be found by setting $d\hat{\rho}/dt = 0$ in equation~\ref{eqn:Master} and solving for $\hat{\rho}$.

\subsection{Measurable Quantities}

Once we have the values of the density matrix elements in the steady state, we would like to be able to extract useful measurable quantities to compare to experiment. 
The response of the medium to an applied field is characterised by the electric susceptibility $\chi$. This macroscopic response of the system can be related to the individual atomic response as
\begin{equation}
    \chi = \frac{-2ND_{ji}^2}{\epsilon_0 \hbar\Omega_{ij}}\rho_{ji}
\end{equation}
where $N$ is the number density (atoms per unit volume in the ensemble), $D_{ji} = \langle j|D|i\rangle$ is the dipole matrix element defined earlier, $\Omega_{ij}$ is the Rabi frequency of the field coupling the $i^{\rm{th}}$ and $j^{\rm{th}}$ levels and $\rho_{ji}$ is the element of the density matrix describing the coherence between the $i^{\rm{th}}$ and $j^{\rm{th}}$ levels. 

The real part of the complex susceptibility is proportional to the dispersion of the medium, while the imaginary part is proportional to the absorption coefficient $\alpha$. 
Often the experimental quantity of interest is the amount of absorption of a laser after passing through the medium, most commonly the absorption of the probe laser which couples the lowest two levels. 
From the above relations we can see that the probe absorption $\alpha_{12}\propto-\Im[\rho_{21}]$, so often we will be interested in the steady state value of the density matrix element $\rho_{21}$.

\subsection{Three-Level Systems}
\begin{figure}
    \centering
    \includegraphics[width = \textwidth]{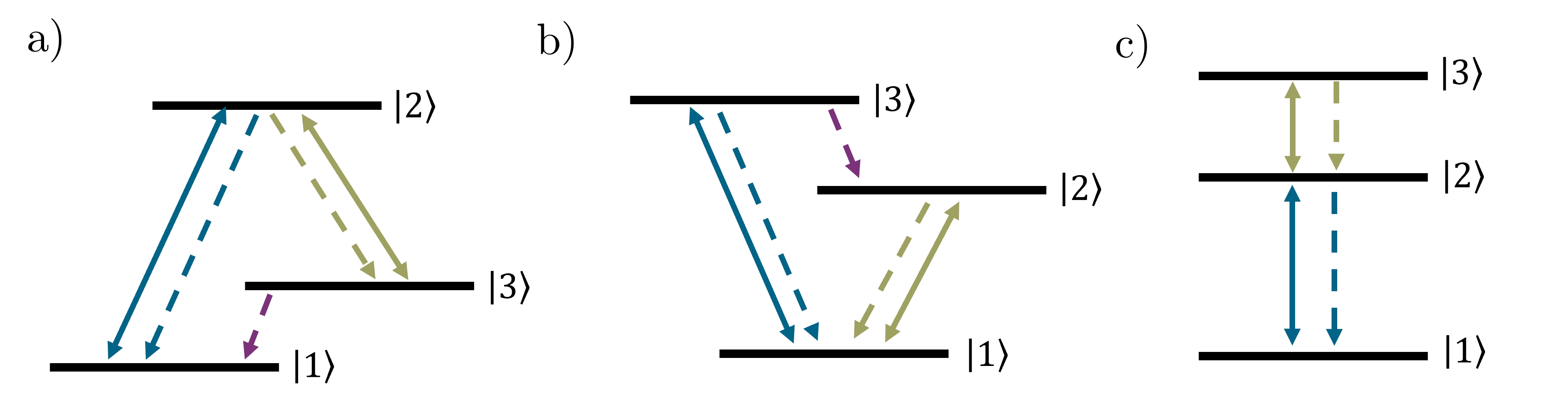}
    \caption{Different possible configurations of 3 atomic levels and 2 driving fields. \newline a) Lambda ($\Lambda)$ b) Vee (V) c) Ladder ($\Xi$) configurations. The solid lines show laser fields coupling two states, the dashed lines show possible decay pathways.}
    \label{fig:3lvl_Configs}
\end{figure}

When considering the simple case of a two-level system we did not have to pay attention to the configuration of the states, however as we add more states and fields coupling these states we need to think about how they are connected. 
For many-level systems there are multiple possible arrangements of energy levels, driving fields and decay pathways within an atom. For 3-level systems, three of the most commonly considered configurations are the Lambda ($\Lambda$), Vee (V) and ladder ($\Xi$) systems, shown in Figure~\ref{fig:3lvl_Configs} from left to right. 
Here we will deal only with ladder systems, in which each level is only coupled to the levels directly above and below it in energy. This level configuration is common in experiments which excite atoms to high-lying Rydberg states through multi-photon excitation \cite{Wade17, Carr12b, Downes20, chen2022}. Obviously this simplified description means that hyperfine structure is not accounted for, but this is often not resolved in higher-lying states and so can be neglected.
%
\begin{figure}
    \centering
    \includegraphics[width = 0.9\textwidth]{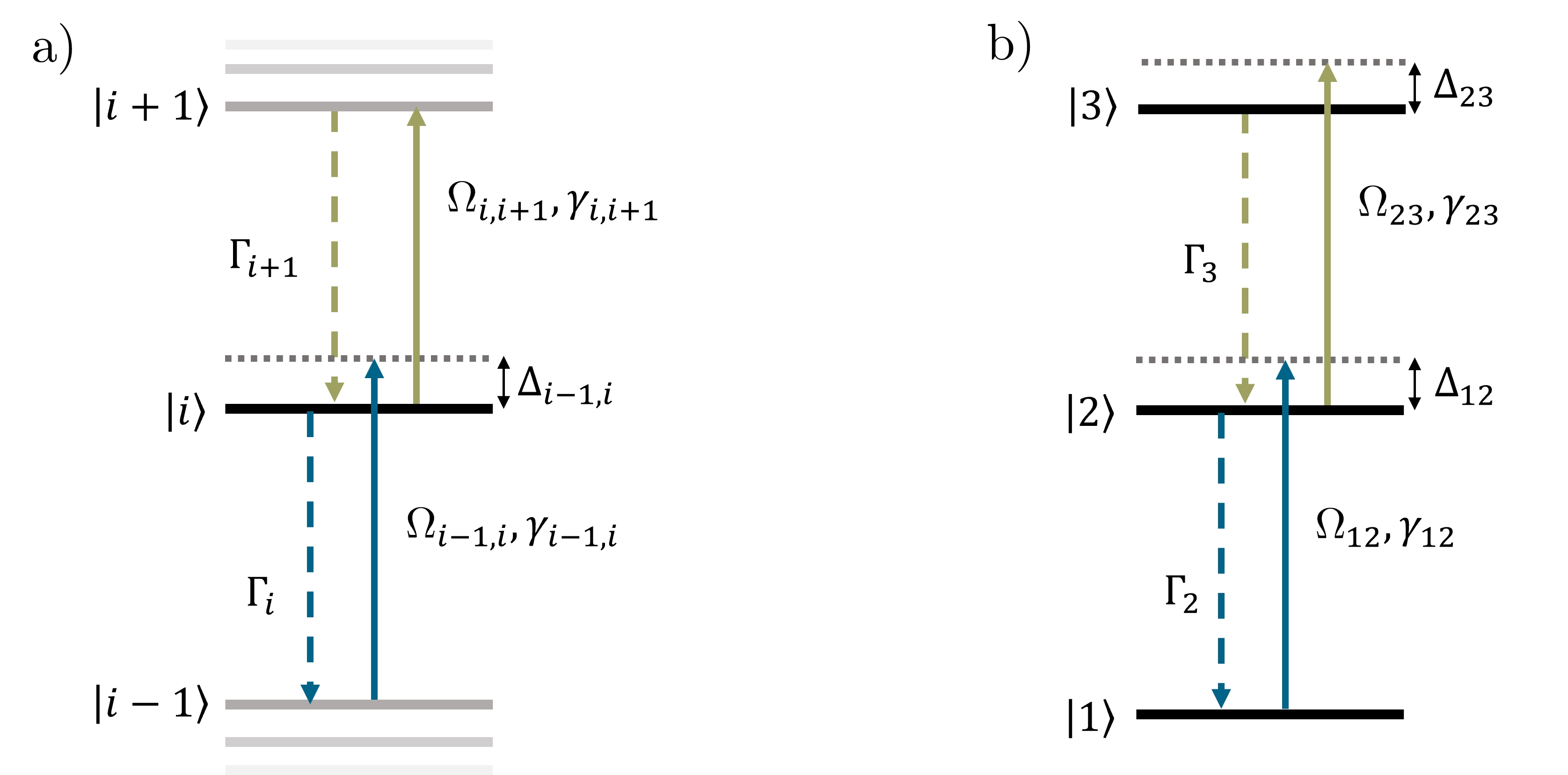}
    \caption{Level diagrams for ladder systems. a) Any level $|i\rangle$ within the ladder will have a field ($\Omega_{i,i+1}, \gamma_{i, i+1}, \Delta_{i,i+1}$) coupling it to the level $|i+1\rangle$ above. There will also be a field ($\Omega_{i-1,i}, \gamma_{i-1,i}, \Delta_{i-1,i}$) coupling to the level $|i-1\rangle$ below. There will also be decay $\Gamma_i$ to the $i-1^{\rm{th}}$ level below. b) A specific example of a 3-level ladder scheme. There is no decay from the lowest energy level ($\Gamma_1 = 0\,\rm{MHz}$).}
    \label{fig:ladder_diags}
\end{figure}

We can then extend our simple two-level model from the previous section to include an extra state $|3\rangle$ and driving field $\Omega_{23}$ that couples the intermediate and highest energy states. A diagram of this simple 3-level ladder scheme and the relevant parameters can be seen in Figure~\ref{fig:ladder_diags}. 
For this 3-level system we apply the same methods and approximations as before and can write the total Hamiltonian in the interaction picture as
\begin{equation}
H = \frac{\hbar}{2}
\begin{pmatrix}
0 & \Omega_{12} & 0 \\
\Omega_{12} & -2\Delta_{12} & \Omega_{23} \\
0 & \Omega_{23} & -2(\Delta_{12} + \Delta_{23})
\end{pmatrix},
\label{eqn:H_3lvl}
\end{equation}
where $\Omega_{12, 23}$ are the Rabi frequencies of the laser fields and $\Delta_{12, 23}$ are the corresponding detunings. 

In the ladder configuration we assume that each level decays to the level directly below it at a rate $\Gamma_n$, and that there is no decay out of the lowest energy level. In this case the diagonal elements of the decay operator $\hat{L}_{\rm{decay}}$ will be given by $L_{ii} = \Gamma_{i+1}\rho_{i+1, i+1} - \Gamma_{i}\rho_{ii}$ while the the off-diagonal elements will be given by $L_{ij,\, i\neq j} = -\frac{\Gamma_i + \Gamma_j}{2}\rho_{ij}$ \cite{Cohen92}. 
This means that for our example 3-level system we have
\begin{equation}
\hat{L}_{\rm{decay}} = \begin{pmatrix} 
\Gamma_2 \rho_{22} & - \frac{\Gamma_2}{2}\rho_{12} & -\frac{\Gamma_3}{2}\rho_{13} \\ 
-\frac{\Gamma_2}{2}\rho_{21} & \Gamma_3 \rho_{33} - \Gamma_2\rho_{22} & -\frac{\Gamma_2 + \Gamma_3}{2}\rho_{23} \\
-\frac{\Gamma_3}{2}\rho_{31} & -\frac{\Gamma_3 + \Gamma_2}{2}\rho_{32} &  - \Gamma_3\rho_{33} \end{pmatrix},
\end{equation}
where we have already made the assumption that there is no decay out of the lowest level ($\Gamma_1 = 0$). 
The elements of the dephasing matrix $\hat{L}_{\rm{dephasing}}$ can be expressed as
\begin{equation}
    L_{ij, i\neq j} = -\left( \gamma_{i,i+1} + \dots + \gamma_{j-1, j}\right) \rho_{ij},
\end{equation}
where $\gamma_{n,n+1}$ is the linewidth of the field coupling the $n^{\rm{th}}$ and $(n+1)^{\rm{th}}$ levels~\cite{GeaBanacloche95}.

\subsection{More Levels}
In Rydberg experiments it is often necessary to use 3 or more fields to reach the desired Rydberg state \cite{Wade17, Carr12b, Downes20, chen2022}. If another field is introduced, for example to couple to adjacent Rydberg states when performing electrometry experiments \cite{Sedlacek12, Holloway14, Holloway14b}, then this again adds to the complexity of the system. These multi-level systems can often be thought of as ladder systems. 
Note that the codes could all be adapted to the other system configurations shown in Figure~\ref{fig:3lvl_Configs}, but the form of the Hamiltonian and decay/dephasing operators would need to be redefined.

\section{Solution Methods}
Even for relatively small numbers of atomic levels ($n>2$) the system of equations for the elements of the density matrix can only be solved numerically unless some assumptions are made, the most common of which is the assumption that the probe field (the field coupling the two lowest energy states) is weak. In this case a perturbative approach can be taken and analytic solutions found (see Section~\ref{sec:anal_sols}), but this assumption is not always valid. Hence it is useful to develop an approach that does not rely on making further assumptions about the system. In this section we will outline a numerical method that does not require the weak-probe assumption, and an analytic solution that is only valid in the weak-probe regime.

\subsection{Matrix Solution}
Any linear system of equations can be expressed as a matrix of coefficients multiplied by a vector of variables. Our aim is to find a way to write the optical Bloch equations in this way, such that we can find the steady state solution using linear algebra. 

\subsubsection{Setting up the Matrices}

First we set up a function to create the density matrix $\hat{\rho}$ itself. 
Given a number of levels, this function returns a matrix populated with \mintinline{python}{Sympy} symbolic objects which will allow us to set up the equations and manipulate the terms.
For an $n$-level system the density matrix will have $n^2$ components. 

\begin{minted}[bgcolor = background]{python}
def create_rho_matrix(levels = 3):
    rho_matrix = numpy.empty((levels, levels), dtype = 'object')
    for i in range(levels):
        for j in range(levels):
            globals()['rho_'+str(i+1)+str(j+1)] = sympy.Symbol(\
            'rho_'+str(i+1)+str(j+1))
            rho_matrix[i,j] = globals()['rho_'+str(i+1)+str(j+1)]
    return numpy.matrix(rho_matrix)
\end{minted}

Next we set up the Hamiltonian of the interaction between the atom and the light, in the form shown in equations~\ref{eqn:H_2lvl} and \ref{eqn:H_3lvl}.  

\begin{minted}[bgcolor = background]{python}
def Hamiltonian(Omegas, Deltas):
    levels = len(Omegas)+1 #count levels
    H = numpy.zeros((levels,levels))
    for i in range(levels):
        for j in range(levels):
            if numpy.logical_and(i==j, i!=0):
                H[i,j] = -2*(numpy.sum(Deltas[:i]))
            elif numpy.abs(i-j) == 1:
                H[i,j] = Omegas[numpy.min([i,j])]
    return numpy.matrix(H/2)
\end{minted}
This function returns a matrix of floats. 
Note that we have omitted the factor of $\hbar$ when we create this Hamiltonian. This is because it will cancel with the factor of $\hbar$ in the Master equation later on. We could include it explicitly, but because it is often many orders of magnitude smaller than the values of $\Omega$ and $\Delta$ it makes the overall values within the Hamiltonian very small and we risk errors due to floating point precision.

Next we set up the matrix form of the decay operator $\hat{\mathcal{L}}$. As we did earlier, we will split this into two separate parts for simplicity. 
First we focus on $\hat{L}_{\rm{decay}}$ describing spontaneous decay. 
The function takes a list of values for the decay rates of the excited states $\Gamma_i$. For an $n$-level system, this list will have $n-1$ entries as we always assume no decay from the lowest energy level ($\Gamma_1 = 0$).
Again the function returns an $n\times n$ matrix populated by multiples of \mintinline{python}{Sympy} symbolic objects. 
\begin{minted}[bgcolor = background]{python}
def L_decay(Gammas):
    levels = len(Gammas)+1
    rhos = create_rho_matrix(levels = levels)
    Gammas_all = [0] + Gammas
    decay_matrix = numpy.zeros((levels, levels), dtype = 'object')
    for i in range(levels):
        for j in range(levels):
            if i != j:
                decay_matrix[i,j] = -0.5*(Gammas_all[i]+\
                Gammas_all[j])*rhos[i,j]
            elif i != levels - 1:
                into = Gammas_all[i+1]*rhos[1+i, j+1]
                outof = Gammas_all[i]*rhos[i, j]
                decay_matrix[i,j] = into - outof
            else:
                outof = Gammas_all[i]*rhos[i, j]
                decay_matrix[i,j] = - outof
    return numpy.matrix(decay_matrix)
\end{minted}
We can create the matrix $\hat{L}_{\rm{dephasing}}$ in a similar way, again from a list of values of $\gamma_{ij}$.
\begin{minted}[bgcolor = background]{python}
def L_dephasing(gammas):
    levels = len(gammas)+1
    rhos = create_rho_matrix(levels = levels)
    deph_matrix = numpy.zeros((levels, levels), dtype = 'object')
    for i in range(levels):
        for j in range(levels):
            if i != j:
                deph_matrix[i,j] = -(numpy.sum(gammas[\
                numpy.min([i,j]):numpy.max([i,j])]))*rhos[i,j]
    return numpy.matrix(deph_matrix)
\end{minted}

Now we have expressions for the Hamiltonian and Lindblad operators, we can put them into the Master equation to get an expression for the time evolution of the density matrix. As we saw in equation~\ref{eqn:Master} we have that 
\begin{equation}
\frac{\partial\hat{\rho}}{\partial t} = -\frac{i}{\hbar}(\hat{H}\hat{\rho} - \hat{\rho}\hat{H}) + \hat{L}_{\rm{decay}} + \hat{L}_{\rm{dephasing}}.
\end{equation}
[Note that the $\hbar$ omitted earlier from the Hamiltonian would have cancelled with the $\hbar$ here, so when we define the function we omit the factor of $1/\hbar$.]

\begin{minted}[bgcolor = background]{python}
def Master_eqn(H_tot, L):
    levels = H_tot.shape[0]
    dens_mat = create_rho_matrix(levels = levels)
    # return Master equation in matrix form
    return -1j*(H_tot*dens_mat - dens_mat*H_tot) + L
\end{minted}

This gives us an expression for the time evolution of each of the components of the density matrix in terms of other components. To make this complex system of equations more convenient to solve we want to write them in the form 
\begin{equation}
\frac{\partial\hat{\rho}_{\rm{vect}}}{\partial t} = \hat{M}\hat{\rho}_{\rm{vect}}
\end{equation}
where $\hat{\rho}_{\rm{vect}}$ is a column vector containing all of the elements of $\hat{\rho}$, and $\hat{M}$ is a matrix of coefficients. 
First we set up a function to create $\hat{\rho}_{\rm{vect}}$, a list of the components of the density matrix $\rho_{i,j}$. 

\begin{minted}{python}
def create_rho_list(levels = 3):
    rho_list = []
    for i in range(levels):
        for j in range(levels):
            globals()['rho_'+str(i+1)+str(j+1)] = sympy.Symbol(\
            'rho_'+str(i+1)+str(j+1))
            rho_list.append(globals()['rho_'+str(i+1)+str(j+1)])
    return rho_list
\end{minted}
We can then create the matrix of coefficients $\hat{M}$ by using the Sympy functions \mintinline{python}{expand} and \mintinline{python}{coeff}. 
The \mintinline{python}{expand} function collects together terms containing the same symbolic object. The \mintinline{python}{coeff} function allows us to extract the multiplying coefficient for a specified symbolic object. This works since the only terms we have are linear with respect to the density matrix elements, e.g. we never have terms in $\rho_{ij}^2$ or higher.

\begin{minted}[bgcolor = background]{python}
def OBE_matrix(Master_matrix):
    levels = Master_matrix.shape[0]
    rho_vector = create_rho_list(levels = levels)
    coeff_matrix = numpy.zeros((levels**2, levels**2), \
    dtype = 'complex')
    count = 0
    for i in range(levels):
        for j in range(levels):
            entry = Master_matrix[i,j]
            expanded = Sympy.expand(entry)
            for n,r in enumerate(rho_vector):
                coeff_matrix[count, n] = complex(expanded.coeff(r))
            count += 1
    return coeff_matrix
\end{minted}

\subsubsection{Time-Dependent Solutions}
We now have an expression for $\partial{\hat{\rho}}_{\rm{vect}}/\partial t$ in terms of the matrix $\hat{M}$ multiplied by $\hat{\rho}_{\rm{vect}}$.
The value of $\hat{\rho}_{\rm{vect}}$ at any time $t$ can be found from
\begin{equation}
    \hat{\rho}_{\rm{vect}}(t) = \exp\left(\hat{M}t\right)\hat{\rho}_{\rm{vect}}(0)
\end{equation}
where $\hat{\rho}_{\rm{vect}}(0)$ is the value of $\hat{\rho}_{\rm{vect}}$ at time $t=0$.
Since we already have a way of creating the matrix $\hat{M}$, it is trivial to define functions to find the time-dependent behaviour of $\hat{\rho}_{\rm{vect}}$.

\begin{minted}[bgcolor = background]{python}
import scipy.linalg as linalg

def time_evolve(operator, t, p_0):
    exponent = operator*t
    #scipy.linalg.expm does matrix exponentiation
    term = linalg.expm(exponent) 
    return numpy.matmul(term, p_0)
\end{minted}

\subsubsection{Singular Value Decomposition}

Often in experiments the measurement timescales are long such that transient oscillatory behaviour such as Rabi oscillations are not observed, and the system reaches a so-called steady-state. 
In the steady state, $\partial{\hat{\rho}}_{\rm{vect}}/\partial t = 0$ so we need to solve the expression 
\begin{equation}
\hat{M}\hat{\rho}_{\rm{vect}} = 0.
\end{equation}

This can be solved numerically in many different ways\footnote{There are built-in routines in numpy/scipy that solve systems of linear equations, such as \mintinline{python}{numpy.linalg.lstsq}. These are based on the SVD but it is tricky to coax out the non-trivial solution. They are also only fractionally faster, as shown by the open blue points in Figure~\ref{fig:comp_time}.}, but here we will solve the system by performing a singular value decomposition (SVD) on the matrix $\hat{M}$ \cite{RHB}. The full details of the theory of the SVD are beyond this discussion, it is suffice to say that the SVD transforms $\hat{M}$ into three matrices such that 
\begin{equation}
\hat{M} = U\Sigma V^{\dag}
\end{equation}
where $V^{\dag}$ is the conjugate transpose of $V$.
The matrix $\Sigma$ is diagonal, the elements corresponding to the singular values of $\hat{M}$. The solution to the system of equations is then the column of $V$ corresponding to the zero singular value. Since we have the same number of equations as we have unknowns we expect only one non-trivial solution. If none of the singular values are zero ($\hat{M}$ is non-singular) then there is no non-trivial solution. Floating point precision means that it is unlikely that any of the values will be identically zero, so we set a limit and assume that values smaller than this are due to floating point errors. 

We perform the SVD using the built-in \mintinline{python}{numpy.linalg.svd} routine which returns the matrices $U$ and $V^{\dag}$ along with an array of singular values $\Sigma$. Before returning the values of the elements of the density matrix we normalise the sum of the populations to be 1. The function returns an array of values for $\hat{\rho}_{\rm{vect}}$, the order of which is the same as the elements of the output of the \mintinline{python}{create_rho_list} function.

\begin{minted}[bgcolor = background]{python}
def SVD(coeff_matrix):
    levels = int(numpy.sqrt(coeff_matrix.shape[0]))
    u,sig,v = numpy.linalg.svd(coeff_matrix)
    abs_sig = numpy.abs(sig)
    minval = numpy.min(abs_sig)
    if minval>1e-12: # to account for floating point precision
        print('ERROR - Matrix is non-singular')
        return numpy.zeros((levels**2))
    index = abs_sig.argmin()
    rho = numpy.conjugate(v[index,:])
    # extract populations and normalise
    pops = numpy.zeros((levels)) 
    for l in range(levels):
        pops[l] = numpy.real(rho[(l*levels)+l])
    t = 1/(numpy.sum(pops))
    rho_norm = rho*t.item() #item() gets Python scalar from array
    return rho_norm
\end{minted}
For ease of calculation we combine all of the previous functions into one function that takes lists of the different parameters ($\Omega$, $\Delta$, $\Gamma$ and $\gamma$) and outputs the steady state solution for the density matrix in the order that \mintinline{python}{rho_list} is created. Note that it outputs the solution of the entire density matrix, so we need to extract the element of interest (usually $\rho_{21}$).

\begin{minted}[bgcolor = background]{python}
def steady_state_soln(Omegas, Deltas, Gammas, gammas = []):
    L_atom = L_decay(Gammas)
    # include dephasing if gammas values given
    if len(gammas) != 0: 
        L_laser = L_dephasing(gammas)
        L_tot = L_atom + L_laser
    else:
        L_tot = L_atom
    H = Hamiltonian(Omegas, Deltas) # build Hamiltonian
    Master = Master_eqn(H, L_tot) # insert H into Master equation
    rho_coeffs = OBE_matrix(Master) # create matrix of coefficients
    soln = SVD(rho_coeffs) # perform SVD to get steady state solution
    return soln
\end{minted}
All we need to do now is pass lists of the system parameters to this function.

\subsection{Analytic Solutions}
\label{sec:anal_sols}
Obviously because the matrix method requires solving an eigenvalue problem for each set of parameters it is not fast, however it does allow for the computation of results without any further assumptions about the system parameters. 
If the system is in the weak-probe regime, defined as the regime where $\Omega_{12}^2/\Gamma_3(\Gamma_2 + \Gamma_3)\ll 1$, a number of assumptions can be made which allow analytic expressions for the probe coherence to be found. Since these assumptions involve treating the probe field as a weak perturbation, we keep only terms linear in $\Omega_{12}$. In this case it can be said that no population transfer occurs, so we have that
\begin{equation}
    \rho_{ii} = 
    \begin{cases}
    1,& i=1 \\
    0,& i>1.
    \end{cases}
\end{equation}
 
For a 3-level system, the probe coherence in the weak-probe regime can be expressed as
\begin{equation}
    \rho_{21} = \frac{-i\Omega_{12}}{2}\left[-i\Delta_{12} +\frac{\Gamma_2}{2}+\gamma_{12}+\frac{\Omega_{23}^2}{4}\left[-i(\Delta_{12} + \Delta_{23})+\frac{\Gamma_3}{2} + \gamma_{12} + \gamma_{23}\right]^{-1}\right]^{-1}.
    \label{eqn:anal_3lvl}
\end{equation}
which can easily be encapsulated into a function. 

\subsubsection{Analytic solution for an arbitrary number of levels} 

It is possible to formulate an expression for the analytic solution in the weak-probe regime for a ladder scheme with any number of levels. This has to be done iteratively. 

For an system with $n$ levels (hence $n-1$ driving fields) we have an expression for the probe coherence which needs to be evaluated for all fields $m=1, 2 \dots n-2, n-1$
\begin{equation}
\rho_{21} = \frac{2iK_1}{\Omega_{12}}
\end{equation}
where
\begin{equation}
K_m = \left(\frac{\Omega_{m, m+1}}{2}\right)^2\left[Z_m + K_{m+1}\right]^{-1}
\end{equation}
and 
\begin{equation} 
Z_m = \sum_{k=1}^m \left(i\Delta_{k,k+1} - \gamma_{k,k+1}\right) -\frac{\Gamma_{m+1}}{2}.
\end{equation}
For example, if we have a 3-level scheme the expression will be
\begin{equation} 
\rho_{21} = \frac{2i}{\Omega_{12}}\left(\frac{\Omega_{12}}{2}\right)^2\left[Z_1 + \left(\frac{\Omega_{23}}{2}\right)^2\left[Z_2\right]^{-1}\right]^{-1}
\end{equation}
where
\begin{equation*}
Z_1 = i\Delta_{12} - \gamma_{12} - \Gamma_2/2
\end{equation*}
and
\begin{equation*}
    Z_2 = i(\Delta_{12} + \Delta_{23}) - (\gamma_{12} + \gamma_{23}) - \Gamma_3/2,
\end{equation*}
which is identical to equation~\ref{eqn:anal_3lvl}.

This has been encapsulated into the function \mintinline{python}{fast_n_level} which takes the same sets of parameters as the \mintinline{python}{steady_state_soln} matrix method and returns the value of $\rho_{21}$.

\begin{minted}{python}
def term_n(n, Deltas, Gammas, gammas = []):
    # Valid for n>0
    if len(gammas) == 0:
        gammas = numpy.zeros((len(Deltas)))
    return 1j*(numpy.sum(Deltas[:n+1])) - \
    (Gammas[n]/2) - numpy.sum(gammas[:n+1])

def fast_n_level(Omegas, Deltas, Gammas, gammas = []):
    n_terms = len(Omegas)
    term_counter = numpy.arange(n_terms)[::-1]
    func = 0
    for n in term_counter:
        prefactor = (Omegas[n]/2)**2
        term = term_n(n, Deltas, Gammas, gammas)
        func = prefactor/(term+func)
    return (2j/Omegas[0])*func
\end{minted}

\subsection{Computation Time}

\begin{figure}[ht]
    \centering
    \includegraphics[width = 1\textwidth]{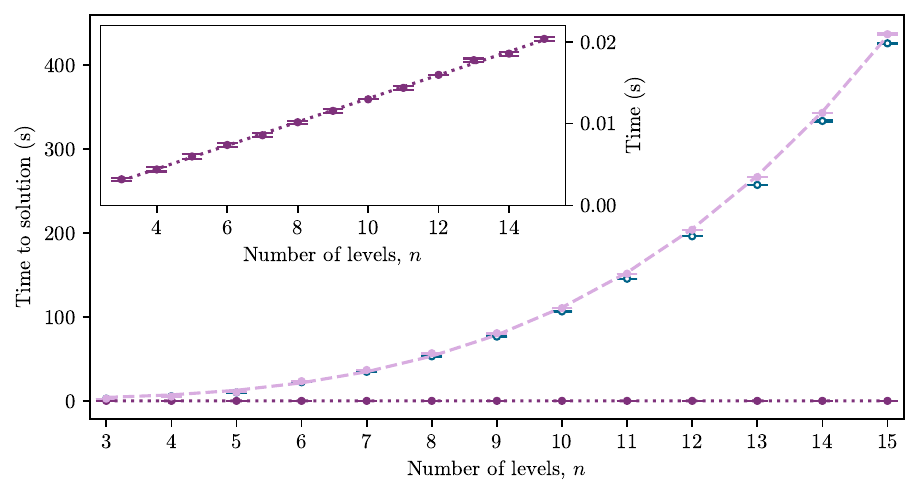}
    \caption{The datapoints show the time taken to find the weak probe coherence for an $n$-level system using the full matrix (light purple) and iterative (dark purple) methods. The errorbars represent the statistical error on 5 repeat calculations using the same hardware. The lines show linear (dark dotted) and power law (light dashed) fits to the iterative and matrix data respectively. Blue open circles show the time taken to use \mintinline{python}{scipy.linalg.lstsq} in place of \mintinline{python}{numpy.linalg.svd} for the matrix method. The inset highlights the linear scaling of the iterative method.}
    \label{fig:comp_time}
\end{figure}

The time taken for both the matrix and iterative solutions to find the weak-probe coherence in an $n$-level system is shown in Figure~\ref{fig:comp_time}. 
The matrix solution is around 3 orders of magnitude slower than the iterative solution, and has a power law dependence on the number of levels. The iterative solution scales linearly with the number of levels, as can be seen in the inset of Figure~\ref{fig:comp_time}.
Obviously the exact times will vary with the hardware used, but it gives an illustration of the usefulness of the iterative method as a tool to fast computation for an arbitrary number of levels. 
Note that none of the code presented here has been optimised for speed. The aim is to get an insight into how the equations are constructed and solved in the simplest way possible, so there are likely much faster but possibly more complex ways to solve the equations presented here.

\section{Examples}
We will now show examples of using both the matrix method and analytic solutions to model the response of a system to applied fields. As noted earlier, often the parameter of interest is the probe coherence $\rho_{21}$ as this is proportional to the absorption of the probe laser through the medium. 
Since $\rho_{21} = \rho_{12}^*$, we will usually extract $\rho_{12}$ from the vector of the solution as this is always at the same index within the vector $\rho_{\rm{vect}}$ (\mintinline{python}{[1]}) whereas the position of $\rho_{21}$ will vary depending on the number of levels. The probe absorption is then $\alpha_{12} \propto \textrm{Im}[\rho_{12}]$.
We will interpret the results by looking at the eigenenergies of the system, found from the total Hamiltonian.

\subsection{3-Level System}
Below is a simple example of using the matrix method (via the \mintinline{python}{steady_state_soln} function) to look at the probe absorption as a function of probe detuning ($\Delta_{12}$) in a 3-level system. We create an array of values for $\Delta_{12}$, then for each of these values we calculate the steady state solution, extract the value of $\rho_{21}$ and store the imaginary part in an array. 

\begin{minted}{python}
Omegas = [0.1,4] # 2pi MHz
Deltas = [0,0] # 2pi MHz
Gammas = [1,0.1] # 2pi MHz

Delta_12 = numpy.linspace(-10,10,200) # 2pi MHz
# create empty array to store solution
probe_abs = numpy.zeros((len(Delta_12))) 
for i, p in enumerate(Delta_12):
    Deltas[0] = p # update value of Delta_12
    # Pass parameters to the function to find the steady state
    # solution for the density matrix
    solution = steady_state_soln(Omegas, Deltas, Gammas)
    # Store the imaginary part of rho_12 in an array
    probe_abs[i] = numpy.imag(solution[1])
\end{minted}

\begin{figure}
\centering
\begin{subfigure}[t]{0.25\linewidth}
\includegraphics[width = \linewidth]{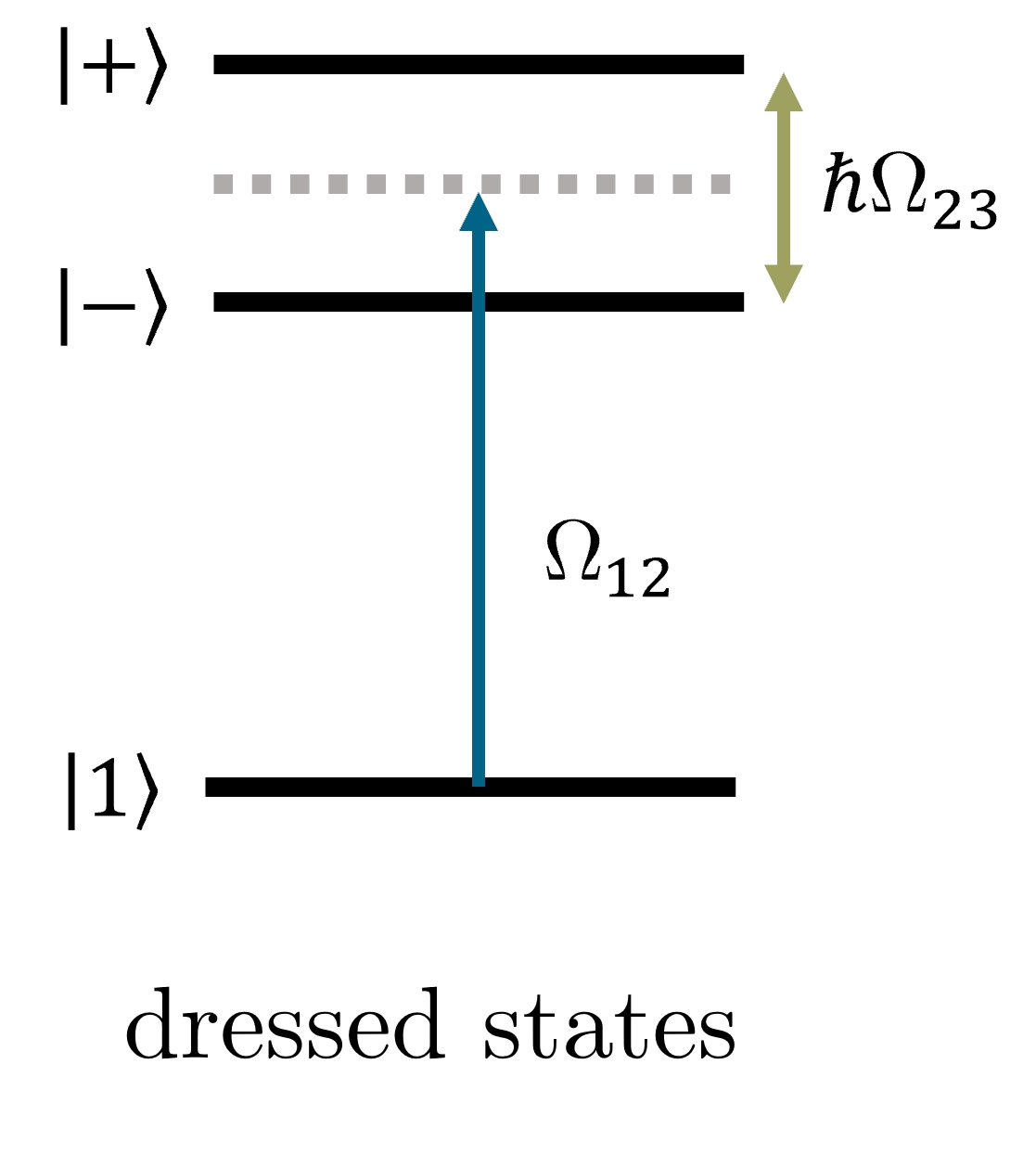}
\end{subfigure}
\hfill
\begin{subfigure}[t]{0.7\textwidth}
\includegraphics[width = \linewidth]{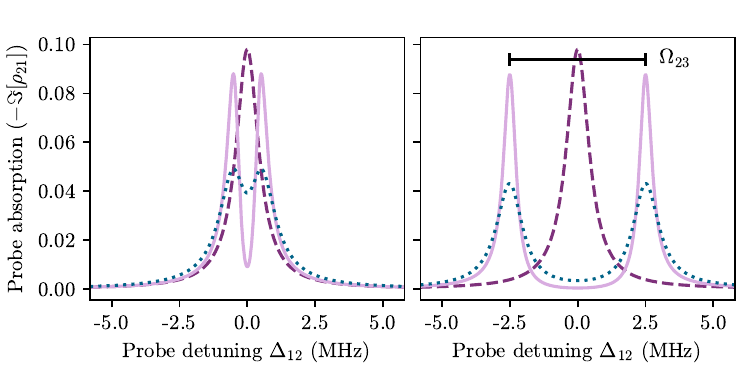}
\end{subfigure}
\caption{Examples of using the matrix method to find the probe laser absorption as a function of detuning for a 3-level system. The response of the medium without the second field ($\Omega_{23} = 0$) is shown by the dark purple dashed line. \textit{Left:} A weak second field ($\Omega_{23} = 1\,\rm{MHz}$) opens a narrow transparency on resonance, as seen by the reduction in the probe absorption at $\Delta_{12} = 0$ on the light purple trace. \textit{Right:} Increasing the Rabi frequency of the second field widens the transparency until the peaks are completely separate. The spacing between the peak maxima is equal to the Rabi frequency of the second field, here $\Omega_{23} = 5\,\rm{MHz}$. In the light purple solutions, dephasing due to laser linewidths has been ignored ($\gamma_{12,23} = 0\,\rm{MHz}$). The dark blue dotted line shows the result of including a small amount of dephasing due to laser linewidth ($\gamma_{12,23} = 0.2\,\rm{MHz}$). This results in the peaks becoming broader and reduced in size. Other parameters remain the same between panels and are set to $\Omega_{12} = 0.1\,\rm{MHz}$, $\Delta_{23} = 0\,\rm{MHz}$, $\Gamma_{2} = 1\,\rm{MHz}$, $\Gamma_{3} = 0.1\,\rm{MHz}$.}
    \label{fig:3_level_varyOmega}
\label{Fig:3level}
\end{figure}

The results of running the code for different parameters are shown in Figure~\ref{fig:3_level_varyOmega}. 
When there is only one field ($\Omega_{23} = 0\,\rm{MHz}$), the system can be thought of as a two-level system. In this case the probe is maximally absorbed on resonance ($\Delta_{12} = 0\,\rm{MHz}$) and we see a Lorentzian absorption feature as a function of probe detuning, the width of which is dependent on the ratio between $\Gamma_2$ and $\Omega_{12}$. 

Setting $\Omega_{12} = 0$ in equation \ref{eqn:H_3lvl} and solving for the 
eigenvectors yields the dressed states
\begin{equation}
|\pm\rangle = (\Delta_{23} \pm\sqrt{{\Delta_{23}}^2 + 
{\Omega_{23}}^2})|2\rangle + \Omega_{23} |3\rangle,
\label{eqn:dressed_states_3lvl}
\end{equation}
with eigenenergies given by
\begin{equation}
E_{\pm} = -\frac{\hbar}{2}\left(2\Delta_{12} + \Delta_{23} \pm \sqrt{\Delta_{23}^2 + \Omega_{23}^2}\right).
\label{eqn:dresseg_eigs_3lvl}
\end{equation}
When both beams are on resonance ($\Delta_{12,23} = 0$) the expression for the dressed states becomes $|\pm\rangle = 1/\sqrt{2}\left(|3\rangle \pm |2\rangle\right)$ and we see that their eigenenergies are separated by $\hbar\Omega_{23}$. 
This means the probe beam that was resonant with the bare state $|2\rangle$ is no longer resonant with either of the dressed states of the coupled atom-field system, so a transparency occurs.  
When the probe becomes resonant with one of the dressed states we see absorption, hence in Figure~\ref{fig:3_level_varyOmega} we see two absorption peaks separated by $\Omega_{23}$. 
This is known as electromagnetically-induced transparency (EIT) or Autler-Townes (AT) splitting, depending on the parameters used. 
The distinction between Autler-Townes splitting and EIT is not well defined but it is generally accepted that EIT requires the condition of weak fields (i.e. small $\Omega_{12,23}$) \cite{Anisimov11}. As $\Omega_{23}$ increases, Autler-Townes splitting becomes the dominant effect and the transparency window widens. 
\begin{figure}
    \centering
    \includegraphics[width=\textwidth]{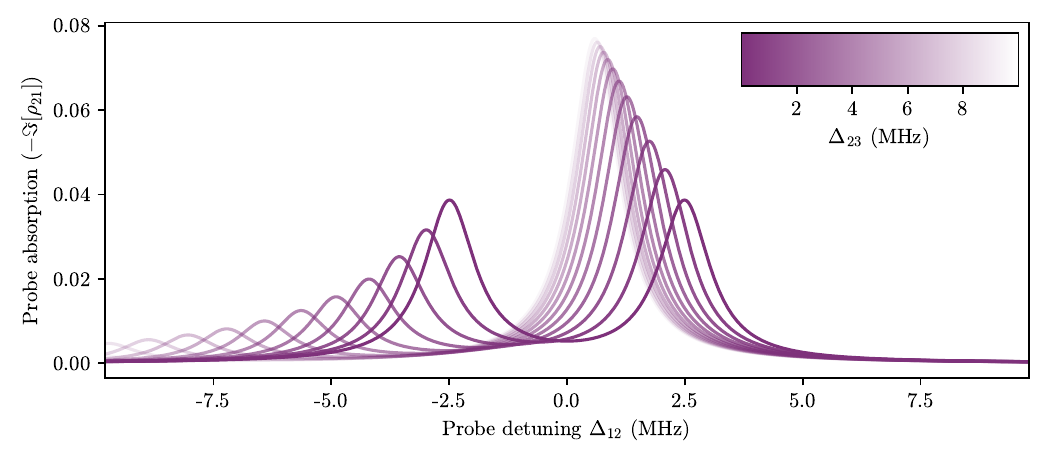}
    \caption{Effect of changing the detuning of the second field ($\Delta_{23}\neq 0$). When $\Delta_{23} = 0$ (darkest line), the two peaks are equal in height since the dressed states contain equal amounts of the bare state $|2\rangle$. They are also symmetrically spaced around $\Delta_{12} = 0$. For $\Delta_{23}>0$ (lighter lines) the peaks become unequal in height since the dressed states contain different amounts of $|2\rangle$. Parameters used are $\Omega_{12} = 0.1\,\rm{MHz}$, $\Omega_{23} = 5\,\rm{MHz}$, $\Gamma_{2,3} = 1\,\rm{MHz}$, $\gamma_{12,23} = 0.1\,\rm{MHz}$.}
    \label{fig:3lvl_varyDelta2}
\end{figure}
For $\Delta_{23} = 0$ both dressed states contain an equal amount of the bare state $|2\rangle$ so the heights of the features in the absorption spectrum will be equal, as seen in Figure~\ref{fig:3_level_varyOmega}. Any detuning of the second field ($\Delta_{23} \neq 0$) will result in an unequal mixture of the bare states in the dressed states as seen in equation~\ref{eqn:dressed_states_3lvl} hence an asymmetric coupling to the ground state. This leads to features of unequal heights in the absorption spectra, as seen in Figure~\ref{fig:3lvl_varyDelta2}. The negatively (positively) detuned peaks are due to coupling with the $|-\rangle$ ($|+\rangle$) dressed state. The peaks also become further apart as $\Delta_{23}$ is increased, since the eigenenergies are separated by $\Delta_{23} + \sqrt{\Delta_{23}^2 + \Omega_{23}^2}$.

\begin{figure}
    \centering
    \includegraphics[width = 1.0\textwidth]{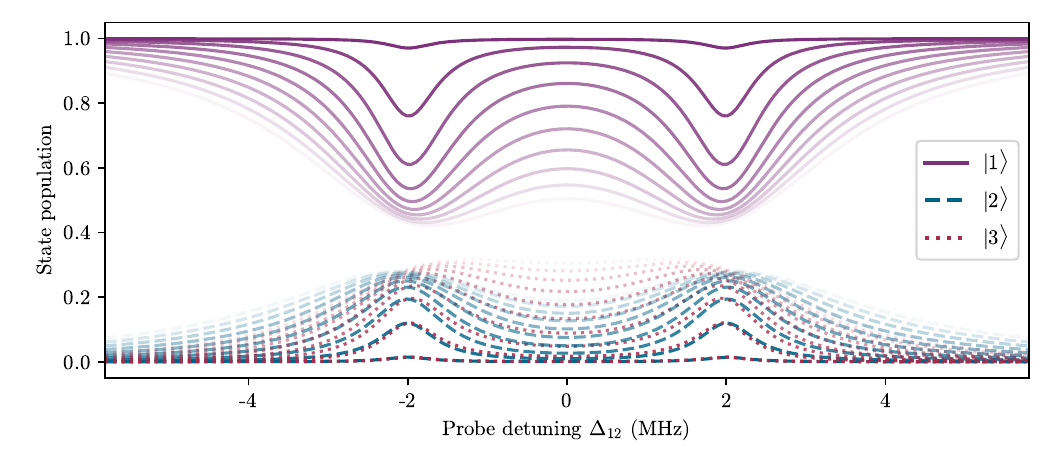}
    \caption{Populations of levels in a 3-level system as a function of probe detuning for different values of probe Rabi frequency. The darkest lines are for $\Omega_{12} = 0.1\,\rm{MHz}$, increasing to $\Omega_{12} = 2\,\rm{MHz}$ with increasing transparency. When the probe beam is weak we see very little population transfer; one of the assumptions made when imposing the `weak-probe regime'. As $\Omega_{12}$ is increased, more population is transferred between states. The other parameters used here are $\Omega_{23} = 4\,\rm{MHz}$, $\Delta_{23} = 0\,\rm{MHz}$, $\Gamma_{2} = 0.5\,\rm{MHz}$, $\Gamma_{3} = 0.2\,\rm{MHz}$ $\gamma_{12,23} = 0.1\,\rm{MHz}$.}
    \label{fig:3_level_pops}
\end{figure}

Using the full matrix solution, it is possible to extract all of the parameters of the density matrix. For example if we wanted to look at the populations of the states we just need to extract the relevant terms from the output. For the 3-level case, the populations of the states $|1\rangle,\,|2\rangle$ and $|3\rangle$ will be the first, fifth and ninth elements respectively. An example of looking at state population as a function of probe detuning and Rabi frequency is shown in Figure~\ref{fig:3_level_pops}. When the probe beam is weak ($\Omega_{12}^2\ll\Gamma_3(\Gamma_2 + \Gamma_3)$) we see very little population transfer; one of the assumptions made when imposing the `weak-probe regime'. As $\Omega_{12}$ is increased (fainter lines in Figure~\ref{fig:3_level_pops}), more population is transferred between states.

\begin{figure}
    \centering
    \includegraphics[width=1\textwidth]{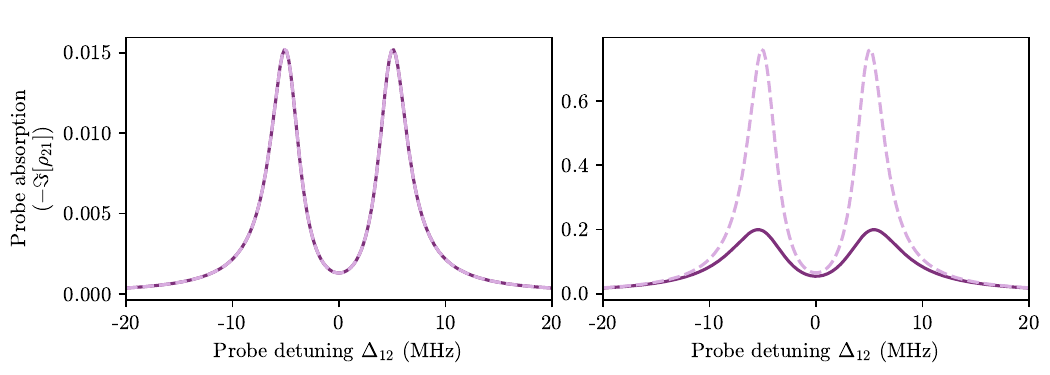}
    \caption{Comparison of results for the matrix (solid dark purple) and analytic weak-probe (dashed light purple) solutions for a 3-level system. \textit{Left}: $\Omega_{12} = 0.1\,\rm{MHz}$ so the weak-probe condition is satisfied and the methods give identical solutions. \textit{Right}: $\Omega_{12} = 5\,\rm{MHz}$ so the weak-probe condition is not satisfied and the methods give very different solutions. Other parameters remain the same and are set at $\Omega_{23} = 10\,\rm{MHz}$, $\Delta_{23} = 0\,\rm{MHz}$, $\Gamma_2 = 5\,\rm{MHz}$, $\Gamma_3 = 1\,\rm{MHz}$, $\gamma_{12,23} = 0.1\,\rm{MHz}$}
    \label{fig:weak_probe_3lvl}
\end{figure}

In Section~\ref{sec:anal_sols} we said that if the weak-probe condition was satisfied ($\Omega_{12}^2\ll\Gamma_3(\Gamma_2 + \Gamma_3)$) it was possible to find analytic solutions for the probe coherence, but that these solutions were not valid outside of the weak-probe regime.
We compare the results of the full matrix solution and analytic solution in Figure~\ref{fig:weak_probe_3lvl}.
In the first case (left hand side of Figure~\ref{fig:weak_probe_3lvl}), the weak-probe approximation is valid (since  $\Omega_{12}^2/\Gamma_3(\Gamma_2 + \Gamma_3) = 1/21 \ll 1$) and so the two methods give equal results. However the analytic solution is many orders of magnitude faster. If the weak-probe condition is not satisfied, the difference in results becomes apparent. This is seen in the right hand side of Figure~\ref{fig:weak_probe_3lvl}. 
The weak-probe condition is no longer satisfied (since  $\Omega_{12}^2/\Gamma_3(\Gamma_2 + \Gamma_3) = 2.3 > 1$) and so the results of the analytic solution no longer match the full matrix solution. 

\subsection{More Levels}

\begin{figure}
    \centering
    \includegraphics[width=1\textwidth]{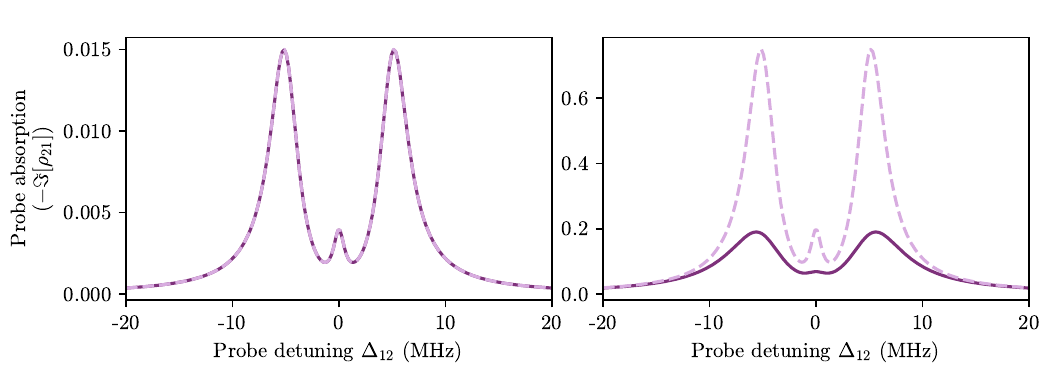}
    \caption{Comparison of results for the matrix (solid dark purple) and analytic weak probe (dashed light purple) solutions for a 4-level system. \textit{Left}: $\Omega_{12} = 0.1\,\rm{MHz}$ so the weak probe condition is satisfied and the methods give identical solutions. 
    \textit{Right}: $\Omega_{12} = 5\,\rm{MHz}$ so the weak probe condition is not satisfied and the methods give very different solutions. Other parameters remain the same and are set at $\Omega_{23} = 10\,\rm{MHz}$, $\Omega_{34} = 2\,\rm{MHz}$, $\Delta_{23,34} = 0\,\rm{MHz}$, $\Gamma_2 = 5\,\rm{MHz}$, $\Gamma_3 = 1\,\rm{MHz}$, $\Gamma_4 = 0.5\,\rm{MHz}$, $\gamma_{12,23,34} = 0.1\,\rm{MHz}$}
    \label{fig:weak_probe_4lvl}
\end{figure}

Now we can look at a 4-level scheme with 3 laser fields ($\Omega_{12}, \Omega_{23}, \Omega_{34}$). 
For the case where $\Omega_{34} = 0\,\rm{MHz}$ we recover the spectrum for the 3-level case as there is no coupling to the $4^{\rm{th}}$ level. As we increase the value of $\Omega_{34}$ a central feature appears and grows while the other features reduce in size.
\begin{figure}
    \centering
    \includegraphics[width = 1.0\textwidth]{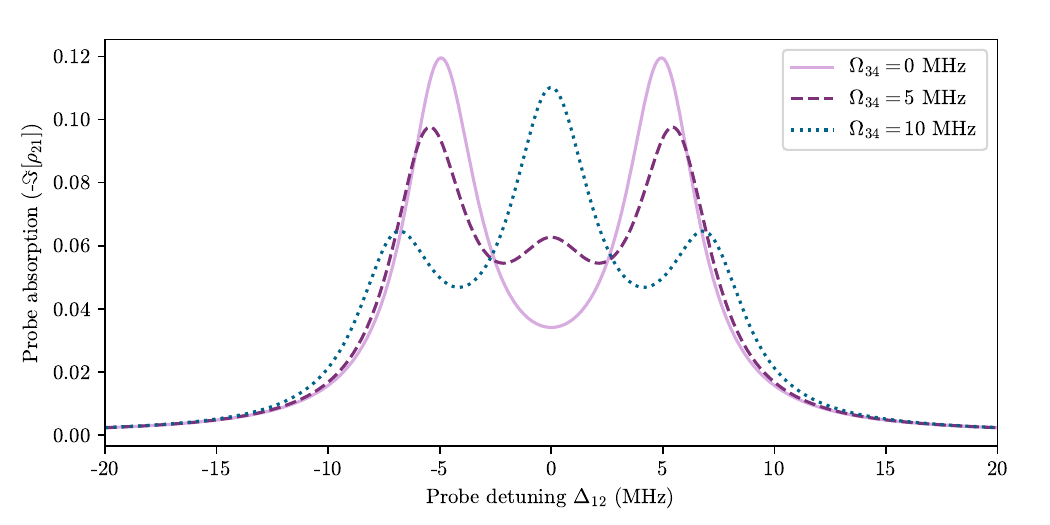}
    \caption{Probe absorption as a function of probe detuning for varying values of $\Omega_{34}$ in a 4-level ladder system. When $\Omega_{34} =0\,\rm{MHz}$ we recover the two Autler-Townes peaks from the 3-level case (solid light purple line). As $\Omega_{34}$ is increased, a central feature begins to appear and the outer peaks reduce in size. Simulation parameters used are: $\Omega_{12} = 1\,\rm{MHz}$, $\Omega_{23} = 10\,\rm{MHz}$, $\Delta_{23,34} = 0\,\rm{MHz}$, $\Gamma_{2,3,4} = 2\,\rm{MHz}$, $\gamma_{12,23,34} = 0.5\,\rm{MHz}$.}
    \label{fig:4_level_a}
\end{figure}
Similarly, for 4 levels we can compare the results of the full matrix solution to the analytic one, as shown in Figure~\ref{fig:weak_probe_4lvl}. 
If we go far outside the weak probe case, then the need for the full matrix solution becomes clear.

\section{Thermal Vapours}
Up until this point we have not considered the effects of temperature on the atomic ensemble. These effects arise due to the motion of the atoms and can include collisional effects and modifications due to interactions with a laser beam of finite size \cite{Finkelstein2023}.  
In this tutorial we will only consider the first-order Doppler effects due to the velocity of the atoms within the ensemble, and we will neglect any collisional or time-of-flight effects. 
We will also assume that all excitation fields are colinear such that we only have to consider the effects of motion in one dimension. Fields propagating in the same direction are referred to as co-propagating, whereas fields propagating in opposite directions are counter-propagating. 
For an atom moving in the same direction as the propagation of a laser field, the frequency of the light will be red-shifted (lower frequency), whereas for an atom moving in the opposite direction the laser frequency would appear blue-shifted. To account for this we can modify the detuning terms in the Hamiltonian as $\Delta_{\rm{eff}} = \Delta - \vec{k}\cdot \vec{v}$ where $\vec{k}$ is the wavevector of the laser light and $\vec{v}$ is the atomic velocity. Hence for an atom moving in the propagation direction we have $\Delta_{\rm{eff}} = \Delta- |k|v$ and for an atom moving opposite to the direction of propagation we get $\Delta_{\rm{eff}} = \Delta + |k|v$. 
\begin{figure}
    \centering
    \includegraphics[width=\textwidth]{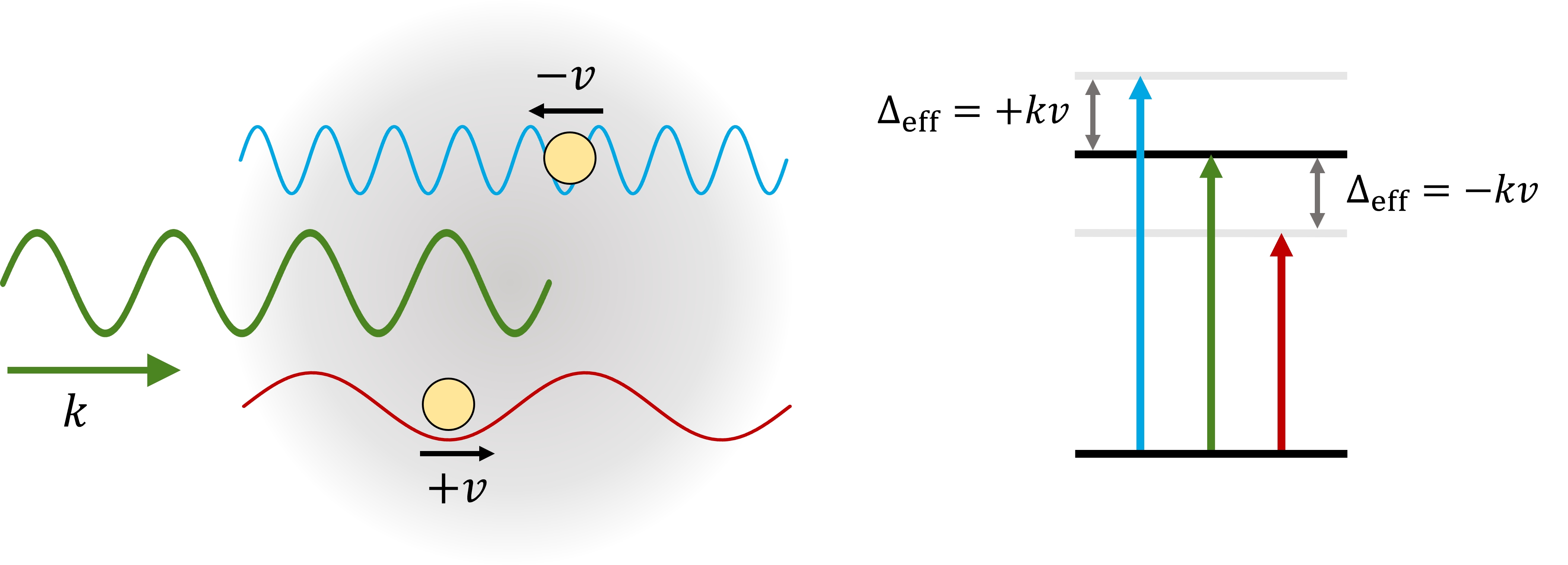}
    \caption{Illustration of how atomic motion affects the frequency `seen' by the atoms. An atom moving at a velocity $v$ opposite to the direction of propagation of a beam with wavevector $k=2\pi/\lambda$ will see a higher (blue-shifted) frequency. Similarly an atom moving at a velocity $v$ with the direction of propagation of the beam will see a lower (red-shifted) frequency. This can be included in the Hamiltonian as an effective detuning, as shown in the level diagram on the right.}
    \label{fig:atom_motion}
\end{figure}

\subsection{Velocity Distribution}
After calculating the response of atoms moving with different velocities, their relative abundances need to be accounted for. 
The atoms will have a distribution of velocities as given by the Maxwell-Boltzmann distribution
\begin{equation}
f_{v}(v_i) = \sqrt\frac{m}{2\pi k_{B}T} \exp \left(-\frac{mv_{i}^2}{2k_{B}T}\right),
\end{equation}
where $m$ is the mass of an atom, $k_B$ is the Boltzmann constant and $T$ is the temperature of the vapour in Kelvin. We can either define this function explicitly, or recognise that this is a Gaussian distribution centred on 0 with width $\sqrt{\frac{k_{B}T}{m}}$ and use a built-in function such as \mintinline{python}{scipy.stats.norm}.

To compute the Doppler averaged response of the vapour we need to take the response of each velocity class and multiply its contribution to the probe absorption by the probability of an atom having that velocity (given by the Maxwell-Boltzmann distribution we defined earlier). Then, for each value of $\Delta_{12}$ we integrate (sum) over all the weighted velocity class contributions.

\subsubsection{Choice of velocity range}

Obviously more velocity classes (a wider range of velocities or finer steps) means more computing time, but one needs to be careful not to lose information. Ideally the simulation would include a wide enough velocity range such that \>99\% of the atoms were accounted for. However for a vapour at room temperature, this would mean including velocities up to $\pm 500 \rm{m/s}$. To get high resolution, this would require a lot of discrete velocity classes. It is possible to truncate without loss of information, but this must be done with intuition about the system. 
Alternatively, one can choose non-linear velocity steps to increase resolution around points of interest. 

\subsection{Example: 3-Level System}

\begin{figure}
    \centering
    \includegraphics[width = \textwidth]{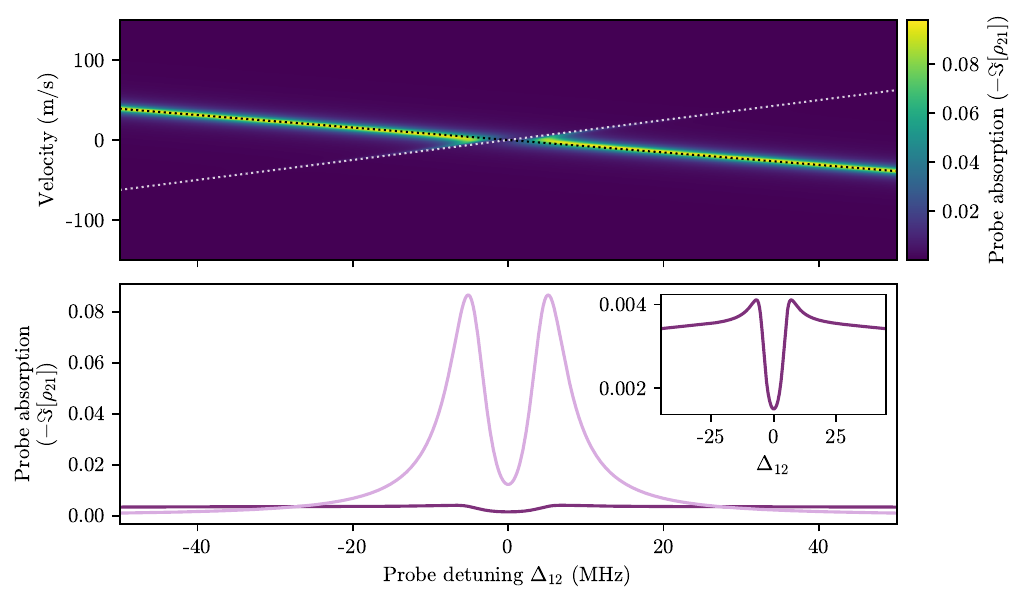}
    \caption{\textit{Top:} Probe absorption as a function of probe detuning and atomic velocity in a 3-level system with counterpropagating beams. The dashed lines represent the eigenvalues of the Hamiltonian in the case of no interaction. \textit{Bottom:} The Doppler-averaged probe absorption (dark purple) found by vertically integrating the response multiplied by the Maxwell-Boltzmann distribution. The light purple shows the response for the zero velocity class for comparison. The shape of the Doppler-averaged response (see inset) is very different to the response for stationary atoms. The parameters are $\Omega_{12}=0.1\,\rm{MHz}$, $\Omega_{23} = 10\,\rm{MHz}$, $\Delta_{23}=0\,\rm{MHz}$, $\Gamma_{2} = 10\,\rm{MHz}$, $\Gamma_{3} = 1\,\rm{MHz}$, $\gamma_{12,23} = 0.1\,\rm{MHz}$, $\lambda_{12} = 780\,\rm{nm}$, $\lambda_{23} = 480\,\rm{nm}$, $s_{12}= +1$, $s_{23} = -1$.}
    \label{fig:Doppler_3lvl}
\end{figure}

In Figure~\ref{fig:Doppler_3lvl} we show an example of calculating the absorption of a weak probe beam in a thermal vapour of 3-level atoms. The relative directions of the beams are taken into account with the \mintinline{python}{direction} variable which can take values of $\pm 1$ (same sign = co-propagating beams, opposite sign = counter-propagating beams). The actual value of the sign does not matter (as the velocity distribution will be symmetric about 0), as long as we are consistent with the signs of the wavevectors relative to each other. 
We also need to specify the wavelengths of the two beams, here chosen to be $\lambda_{12} = 780\,\rm{nm}$ and $\lambda_{23} = 480\,\rm{nm}$ to match commonly used rubidium 2-photon EIT experiments. The beams have been set to be counter-propagating.
These calculations can be quite computationally intensive, especially when using the full matrix method. Since the purpose here is to illustrate the effects of atomic motion, we will work in the weak probe regime ($\Omega_{12}^2 \ll \Gamma_3(\Gamma_2 + \Gamma_3)$) and use the iterative solution method (via the \mintinline{python}{fast_n_level} function) to save time. 
\begin{minted}{python}
Omegas = [0.1,10] # 2pi MHz
Deltas = [0,0] # 2pi MHz
gammas = [0.1,0.1] # 2pi MHz
Gammas = [10,1] # 2pi MHz
lamdas = numpy.asarray([780, 480])*1e-9 # m
directions = numpy.asarray([1,-1]) # unitless direction vectors
ks = directions/lamdas # 2pi m^-1

Delta_12 = numpy.linspace(-50,50,500)  # 2pi MHz
velocities = numpy.linspace(-150, 150, 501) # m/s
probe_abs = numpy.zeros((len(velocities), len(Delta_12)))
for i, v in enumerate(velocities):
    # loop over all atomic velocities
    for j, p in enumerate(Delta_12):
        # loop over all values of probe detuning (lab frame)
        Deltas[0] = p
        # create empty array to store modified detunings
        Deltas_eff = numpy.zeros(len(Deltas))
        for n in range(len(Deltas)):
            Deltas_eff[n] = Deltas[n]+(ks[n]*v)*1e-6 # in MHz
        # pass parameters (including modified detuning) to function
        solution = fast_n_level(Omegas, Deltas_eff, Gammas,\
        gammas)
        probe_abs[i,j] = -numpy.imag(solution)
\end{minted}
The two dotted lines in the upper plot of Figure~\ref{fig:Doppler_3lvl} are the positions at which the dressed states $|+\rangle$ and $|-\rangle$ as defined in equation~\ref{eqn:dresseg_eigs_3lvl} are resonant with the probe. This will occur when 
\begin{equation}
E_{\pm} = -\Delta_{\rm{12_{eff}}} - \frac{\Delta_{\rm{23_{eff}}}}{2} \pm \frac{1}{2}\sqrt{(\Delta_{\rm{23_{eff}}}^2 + \Omega_{23}^2)} = 0,
\end{equation}
where the effective detunings are given by
\begin{align*}
    \Delta_{\rm{12_{eff}}} =&\, \Delta_{12} + v s_{12}k_{12} \\
    \Delta_{\rm{23_{eff}}} =&\, \Delta_{23} + v s_{23}k_{23}.
\end{align*}

$k_{12,23} = 1/\lambda_{12,23}$ are the magniutudes of the wavevectors (ignoring the factor of $2\pi$) and $s_{12,23} = \pm 1$ represent the relative direction of the beams. $E_{\pm}$ are the eigenvalues of the Hamiltonian with $\Omega_{12} = 0$ and detunings replaced by the effective detunings.
In the limit of large detuning where $\Delta_{\rm{23_{eff}}}\gg \Omega_{23}$, the equations for the lines are
\begin{align}
v_+ =& \frac{-\Delta_{12}}{s_{12}k_{12}} \\
v_- =& \frac{-(\Delta_{12}+\Delta_{23})}{s_{12}k_{12} + s_{23}k_{23}}.
\end{align}

If fewer velocity classes were included, the integrated result would look very different and becomes meaningless. We can however use the knowledge of the equations of the asymptotic lines (found via the eigenvalues) to give an idea of the range of velocities that need to be sampled.

\subsection{Example: 4-Level System}

We can use the same method to look at the effects of atomic motion on 4-level atoms. As with the previous case, we consider systems that are within the weak probe limit so we can use the faster iterative solution. Since we are only considering the 1D case we can again specify beam direction using the \mintinline{python}{direction} variable.

\begin{figure}
    \centering
    \includegraphics[width = \textwidth]{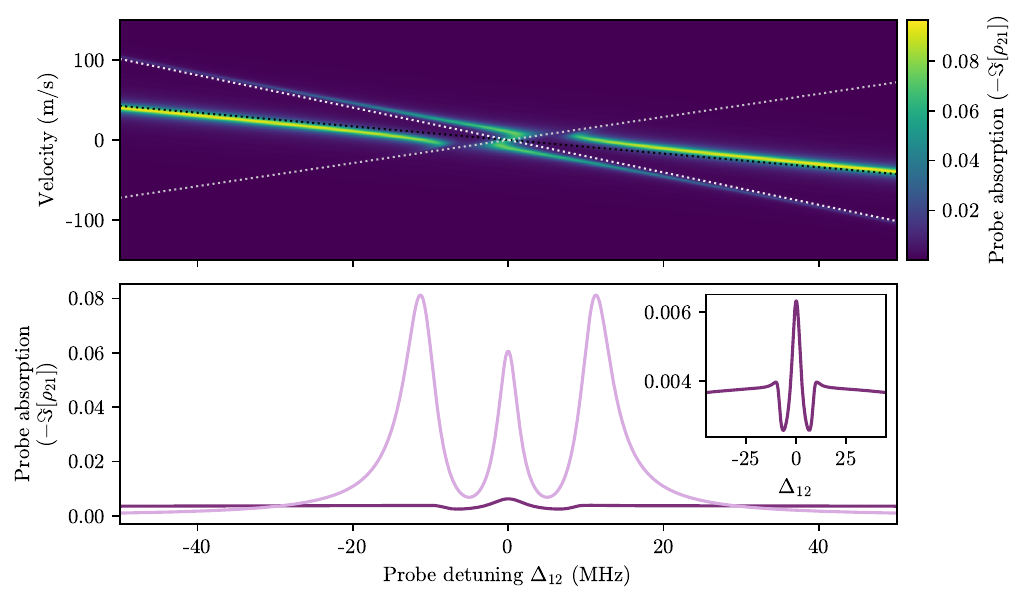}
    \caption{\textit{Top:} Probe absorption as a function of probe detuning and atom velocity in a 4-level system where the probe beam counter-propagates with the other two. The dashed lines represent the eigenvalues of the Hamiltonian in the case of no interaction. \textit{Bottom:} The Doppler-averaged probe absorption (dark purple) found by vertically integrating the response multiplied by the Maxwell-Boltzmann distribution. The light purple shows the response for the zero velocity class for comparison. The shape of the Doppler-averaged response is very different to the response for stationary atoms (see inset).The parameters are $\Omega_{12}=1\,\rm{MHz}$, $\Omega_{23} = 20\,\rm{MHz}$, $\Omega_{34} = 10\,\rm{MHz}$, $\Delta_{23,34}=0\,\rm{MHz}$, $\Gamma_{2} = 10\,\rm{MHz}$, $\Gamma_{3,4} = 1\,\rm{MHz}$, $\gamma_{12,23,34} = 0.1\,\rm{MHz}$, $\lambda_{12} = 852\,\rm{nm}$, $\lambda_{23} = 1470\,\rm{nm}$, $\lambda_{34} = 843\,\rm{nm}$, $s_{12}= +1$, $s_{23,34} = -1$.}
    \label{fig:Doppler_4lvl}
\end{figure}

In the upper plot of Figure~\ref{fig:Doppler_4lvl} we can see 3 asymptotic `branches', whose positions can again be found by finding the eigenvalues of the Hamiltonian in the case where $\Omega\ll\Delta_{\rm{eff}}$, and equating them to zero. Rewriting the expressions in terms of $v$ we have 
\begin{align}
v_1 =& \frac{-\Delta_{12}}{s_{12} k_{12}} \\
v_2 =& \frac{-\Delta_{12}-\Delta_{23}}{s_{12} k_{12} + s_{23} k_{23}} \\
v_3 =& \frac{-\Delta_{12}-\Delta_{23}-\Delta_{34}}{s_{12} k_{12} + s_{23} k_{23} + s_{34} k_{34}},
\end{align}
where again $k_{12,23,34} = 1/\lambda_{12,23,34}$ are the magnitudes of the wavevectors and $s_{12,23,34} = \pm 1$ represent the relative directions of the beams.
Knowing the equations of these asymptotic lines can be useful when deciding how many velocity classes need to be taken into account. In the above example, there is not much extra information gained by including velocity classes $>100\,\rm{m/s}$ or $<-100\,\rm{m/s}$.

\section{Conclusion}
In this tutorial we have outlined methods of solving the optical Bloch equations and demonstrated how they can be implemented simply in Python. We began by looking at how the optical Bloch equations are derived from the semi-classical treatment of atom-light interactions, the role of the density matrix and how to extract measurable quantities from the values.
We presented two solution methods in two different regimes; an analytic solution when the system is in the weak-probe regime and the matrix method which requires fewer simplifying assumptions. 
We then demonstrated the use of these solution methods to examine example spectra for few-level systems in both cold and thermal ensembles. 
All functions presented here are available on GitHub \url{https://github.com/LucyDownes/OBE_Python_Tools}, along with example Jupyter notebooks. The codes used to create all plots in this paper are also available there.

\section*{Acknowledgements}
The author would like to thank Ifan Hughes for constructive comments on an early draft of the manuscript, and Ollie Farley for code checking and GitHub help.

\section*{References}

\bibliographystyle{iopart-num}
\bibliography{OBE_Paper}

\end{document}